\documentclass[aps,pra,twocolumn, showpacs]{revtex4-1}
\usepackage{latexsym,graphicx,graphics,amsmath,amssymb}
\usepackage{diagbox}
\usepackage{array}
\usepackage[table]{xcolor}
\usepackage{booktabs}
\usepackage{amstext}
\usepackage{hyperref}

\newcolumntype{x}[1]{%
>{\centering\hspace{0pt}}p{#1}}%
\begin{document}

%\markboth{Physical Review A}{}

\title{Noise and measurement errors in a practical two-state quantum bit commitment protocol}

\author{Ricardo Loura$^{1,2,\ast}$, \'Alvaro J. Almeida$^{3,4}$, Paulo S. Andr\'e$^{3,4}$, Armando N. Pinto$^{4,5}$, Paulo Mateus$^{1,2}$ and Nikola Paunkovi\'c$^{1,2,}$}
\email{ricardoloura@gmail.com, npaunkov@math.ist.utl.pt}
\affiliation{$^{1}$SQIG -- Instituto de Telecomunica\c{c}\~oes, IST-TU Lisbon, 1049-001 Lisbon, Portugal}
\affiliation{$^{2}$Department of Mathematics, IST, Technical University of Lisbon, 1049-001, Lisbon, Portugal}
\affiliation{$^{3}$Department of Physics, University of Aveiro, 3810-193 Aveiro, Portugal}
\affiliation{$^{4}$Instituto de Telecomunica\c{c}\~oes, University of Aveiro, 3810-193 Aveiro, Portugal}
\affiliation{$^{5}$Department of Electronics, Telecommunications and Informatics, 3810-193 Aveiro, Portugal}

%%%%%%%%%%%%%%%%%%%%%%%%%%%%%%%%%%%%%%%%%%%%%%%%%%%%%%%%%%%%%%%%%%%%%%%%%%%%%%%%%%%%%%%%%%%%%%%%%%%%%%%%%%%%%%%%%%%%%%%%%%%%%%%%%%%%%%%%%%%%%%%%%%%%%%%%%%%%%%%%%%%%%%%%%%
%%%%%%%%%%%%%%%%%%%%%%%%%%%%%%%%%%%%%%%             ABSTRACT             %%%%%%%%%%%%%%%%%%%%%%%%%%%%%%%%%%%%%%%%%%%%%%%%%%%%%%%%%%%%%%%%%%%%%%%%%%%%%%%%%%%%%%%%%%%%%%%%
%%%%%%%%%%%%%%%%%%%%%%%%%%%%%%%%%%%%%%%%%%%%%%%%%%%%%%%%%%%%%%%%%%%%%%%%%%%%%%%%%%%%%%%%%%%%%%%%%%%%%%%%%%%%%%%%%%%%%%%%%%%%%%%%%%%%%%%%%%%%%%%%%%%%%%%%%%%%%%%%%%%%%%%%%%

\begin{abstract}
We present a two-state practical quantum bit commitment protocol, the security of which is based on the current technological limitations, namely the non-existence of either stable long-term quantum memories, or non-demolition measurements. For an optical realization of the protocol, we model the errors, which occur due to the noise and equipment (source, fibers and detectors) imperfections, accumulated during emission, transmission and measurement of photons. The optical part is modeled as a combination of a depolarizing channel (white noise), unitary evolution (e.g. systematic rotation of the polarization axis of photons) and two other basis-dependent channels, the phase- and the bit-flip channels. We analyze quantitatively the effects of noise using two common information-theoretic measures of probability distribution distinguishability: the fidelity and the relative entropy. In particular, we discuss the optimal cheating strategy and show that it is always advantageous for a cheating agent to add some amount of white noise -- the particular effect not being present in standard quantum security protocols. We also analyze the protocol's security when the use of (im)perfect non-demolition measurements and noisy/bounded quantum memories are allowed. Finally, we discuss errors occurring due to a finite detector efficiency, dark counts and imperfect single-photon sources and show to have the same effects as those of standard quantum cryptography.
\end{abstract}


\maketitle%068

\section{Bit commitment}

Among security tasks, the bit commitment protocol holds a prominent role as it represents a computational primitive for many important information processing protocols. It is a two-party protocol that consists of two phases: the {\em commitment} and the {\em opening} phases. In the commitment phase, one client (Alice) {\em commits} to a value of a bit (commits to either $0$ or $1$) at a certain moment in time $t_0$. After performing the commitment, Alice finalizes the protocol by revealing ({\em opening}) her choice to the other client (Bob), at some later moment in time $t_1$. The commitment to a certain value could be seen, for instance, as a promise to either perform a certain action in a future moment in time $t_2 \geq t_1$ (e.g. buy a house from Bob for a given fixed price $X$) -- commitment to $1$, or not -- commitment to $0$. The protocol has to fulfill three security requirements: Alice cannot change her commitment later in time, in particular during the opening phase (the protocol is {\em binding}); Bob cannot learn Alice's commitment before the opening phase (the protocol is {\em concealing}); if both clients are honest (if they execute the protocol according to the rules), then Bob will successfully open Alice's commitment (the protocol is {\em viable}). Commitment schemes are nowadays an important phase on several cryptographic protocols, in particular in {\em zero-knowledge proof systems} and {\em authentication protocols} (for a more detailed description, see Appendix~\ref{bit_commitment_applications}).

The idea behind the classical solutions to this problem is to {\em lock} Alice's commitment in a secure ``safe'' (the commitment phase), such that without the key it is impossible to break into it, and give that ``safe'' to Bob. During the opening phase, Alice gives Bob the key, and he learns her commitment. One way of doing this is to use ordinary keys and locks. Another, to perform the locking by encrypting the commitment, using a secret encryption key. In both cases, the solutions have to meet the binding and concealment requirements. Unfortunately, it is not possible to perform a bit commitment protocol that is unconditionally secure. If Alice chooses to protect her commitment by placing it in a real safe, since there are no unbreakable safes, the protocol would not be unconditionally concealing. If she chooses to protect her commitment by encrypting it using a secret key known only by her, she can achieve the concealing requirement, but then the protocol would no longer be binding. Namely, no  unbreakable encryption scheme is at the same time binding -- whatever commitment value Alice encrypted during the commitment phase, she can always present a suitable key that would decrypt to either of the two commitment values.

Attempts to solve this problem using quantum systems have been done previously~\cite{Brassard_quantum_bit_commitment}, but it was shown that no quantum bit commitment protocol can be both unconditionally binding and concealing~\cite{no_qbit_commitment}. Nevertheless, it is possible to, using the current technological limitations, perform a practical quantum bit commitment protocol that will be secure in a foreseeable future~\cite{Danan}. The protocol is based on the famous BB84 quantum cryptographic protocol~\cite{bb84}. The proposed protocol is binding, due to the unconditional security of the BB84 protocol~\cite{qcryptography-security} and the fact that we do not have long-term stable quantum memories, nor do we have apparatuses able to perform non-demolition measurements of photons. Practical bit commitment protocols whose security is based on limited amount of quantum memories were studied previously~\cite{Bound_Memory}, as well as protocols whose security is based on having imperfect (due to unavoidable noise) quantum memories~\cite{Noisy_Memory}. One such protocol, using entangled states, was recently reported to be experimentally performed~\cite{Ng_2012}. Finally, we note that the above no-go theorem on unconditional security of (quantum) bit commitment schemes is applicable only to non-relativistic protocols. Using relativistic effects, it is possible to design an unconditionally secure bit commitment protocol~\cite{Kent_1999}, which was recently implemented~\cite{relativisticBC}.

In this paper, we present a two-state version of the practical quantum bit commitment protocol based on the B92 cryptographic protocol \cite{b92}. Its practical security relies on the fact that long-term stable quantum memories and non-demolition measurements are currently out of reach. Provided those technological limitations, our protocol is more secure than classical counterparts, as its security is based on physical laws, rather than on computational assumptions. We also study the effects of noise, source imperfections and measurement errors on the protocol's security. In particular, we show that adding a certain amount of white noise always increases the chances of a dishonest Alice to cheat Bob and postpone her decision until the opening phase. Finally, we analyze the security of the protocol in the presence of (im)perfect non-demolition measurements and noisy/bounded quantum memories.

\section{Two-state quantum protocol}
\label{the_protocol}

The protocol is based on quantum complementarity -- the impossibility to simultaneously measure two non-commuting observables. Therefore, one has to decide to measure only one out of two possible observables of a physical system, and obtain information about only one of two features of a system. The choice of measurement can be interpreted as a commitment to a bit value, and the measurement outcome used as a proof of this particular choice (i.e. commitment). This is somewhat the opposite approach to that used in classical solutions: instead of (securely) imprinting the information of a commitment choice into a state of a physical system (writing down an encrypted message on a piece of paper, for example), the choice is done by acquiring information about {\em only one out of two} possible features of a physical system. This approach has already been used for designing quantum contract signing~\cite{qsig-previous,qsig} and simultaneous dense coding protocols~\cite{sim_dense_coding}. In those cases, the security of the protocols is provided by the laws of physics (e.g. quantum mechanics), rather than by the computational complexity of the decryption schemes. Also, a ``probabilistic'' two-state quantum bit string commitment protocol, based on the same mechanism, was recently proposed~\cite{kent}, in which Alice commits to a string of $n$ bits, such that Bob can learn not more than $m<n$ bits, up to negligible probability (note that because this protocol is quantum its unconditional security is not implying the existence of unconditionally secure quantum bit commitment schemes, as would be the case for its classical counterpart~\cite{kent}). For similar work on coin tossing and bit-string generation, see~\cite{Kent_2003, Lamoureux_2005, Barrett_2004, Buhrman_2008}.

In addition to the commitment and the opening phases, the quantum protocol begins with the {\em initialization} phase, during which Bob prepares a number of identical physical two-level systems (qubits) on which Alice is to perform the commitment measurement (the same measurement on each qubit). Bob sends to Alice a number of qubits, each randomly prepared in one of two given quantum states ($|0\rangle$ or $|1\rangle$), without revealing any information about the prepared states. During the commitment phase, Alice chooses only one out of two non-commuting observables, $\hat{C}_0$ or $\hat{C}_1$ (given by the two states used by Bob, see below), measures it on each qubit received from Bob, and keeps the record of measurement outcomes. Finally, during the opening phase, she reveals the results to Bob, which serves as a proof of her commitment.

We require that the qubit states $|0\rangle$ and $|1\rangle$ used in the protocol are not orthogonal~\footnote{{Note that we fix the scalar product to be a real number. As the two states $|0\rangle$ and $|1\rangle$ are used to design Alice's measurement observables $\hat{C}_0$ and $\hat{C}_1$, the relative phase between the two vectors is irrelevant and for simplicity we take it to be zero. Note also that, as the relevant quantity in calculations is $\cos\theta$, for reasons of simplicity we do not use standard Bloch representation of pure qubit states. A simple substitution $\theta \rightarrow \theta /2$ in formulas gives results presented below in standard form of Bloch representation.}}, $\langle 0|1 \rangle = \cos \theta$, with $\theta \in (0,\pi /2)$. Let us denote the states orthogonal to $|0\rangle$ and $|1\rangle$ as $|0^\bot\rangle$ and $|1^\bot\rangle$, respectively: $\langle 0^\bot|0\rangle=0$ and $\langle 1^\bot|1\rangle=0$. This way, we defined two (orthonormal) bases $\mathcal{B}_0 = \{|0\rangle,|0^\bot\rangle \}$ and $\mathcal{B}_1 = \{|1\rangle,|1^\bot\rangle \}$, which in turn define two orthogonal observables $\hat{C}_0$ and $\hat{C}_1$:
\begin{equation}
\begin{aligned}
\label{commitment_obs}
\hat{C}_0 = 0 \! \cdot \! |0\rangle\langle 0 | + 1 \! \cdot \! |0^\bot\rangle\langle 0^\bot|\,,\\
\hat{C}_1 = 1 \! \cdot \! |1\rangle\langle 1 | + 0 \! \cdot \! |1^\bot\rangle\langle 1^\bot|\,.
\end{aligned}
\end{equation}
Finally, we list the eigenvectors of $\hat{C}_1$ expressed~\footnote{{Note that, for reasons of simplicity, we take the particular choice of the gauge for state $|1^\bot \rangle$, fixing the relative phases between it and states $|0 \rangle$ and $|0^\bot \rangle$ to be $0$ and $\phi$, respectively.}} in the $\mathcal{B}_0$ basis:
\begin{equation}
\begin{aligned}
	|1 \rangle & = \cos \theta |0 \rangle + e^{i\phi}\sin \theta |0^\bot \rangle \\
	|1^\bot \rangle & = \sin \theta |0 \rangle - e^{i\phi}\cos \theta |0^\bot \rangle\,,
\end{aligned}
\end{equation}
and vice versa:
\begin{equation}
\begin{aligned}
	|0 \rangle & = \cos \theta |1 \rangle + \sin \theta |1^\bot \rangle \\
	|0^\bot \rangle & = e^{-i\phi}(\sin \theta |1 \rangle - \cos \theta |1^\bot \rangle)\,,
\end{aligned}
\end{equation}
for some $\phi \in [0,2 \pi)$.

We can now give a more detailed description of our two-state practical quantum bit commitment protocol. It consists of three phases, arranged in chronological order ($T_0 < T_1 < T_2$):

{\bf{\em The Initialization Phase:}} At time $T_0$, Bob randomly chooses a string of $N$ bits $(b_1,b_2, \ldots b_N)$, with $b_k \in\{ 0,1 \}$, and sends a string of $N$ qubits to Alice, each prepared in the pure state $|b_k \rangle$, and emitted at time $t_k$, with $k \in \{1,2, \ldots N\}$. Bob keeps the information of the states $|b_k \rangle$ of each qubit, as well as the times $t_k$ of the emission of each qubit. We assume that $t_1 < t_2 < \ldots t_N$ and that $t_N - t_1 << T_2 - T_1$.

{\bf{\em The Commitment Phase:}} At time $T_1$, Alice starts measuring on {\em all} the qubits received {\em only one} observable, either $\hat{C}_0$ or $\hat{C}_1$, depending on her commitment choice ($\hat{C}_0$ corresponds to the commitment to value $0$, $\hat{C}_1$ to value $1$). She announces the arrival times of each qubit (which are at the same time the times of measurement of each qubit; see below for the discussion), a string $(\tau_1,\tau_2, \ldots \tau_n)$, with $\tau_1 = T_1$, and keeps the record of the measurement results to her, a string $(r_1,r_2, \ldots r_n)$, with~\footnote{{Note that some, in fact with todays's technology many, qubits sent will be either lost during their transmission, or not detected due to imperfect detectors.}} $n \leq N$.

{\bf{\em The Opening Phase:}} At time $T_2$, Alice reveals her commitment $c\in \{ 0,1 \}$ (e.g. the measurement observable $\hat{C}_c$), together with the measurement results $r_i$, with $i\in \{ 1, \ldots n \}$, to Bob.

Note that not all qubits sent by Bob arrive to and are measured by Alice. Thus, $n \leq N$, and for each index $i$ labeling Alice's measurement times $\tau_i$ and results $r_i$, there is a corresponding index $k = k(i)$ labeling Bob's qubit emission times $t_{k(i)}$ and corresponding bits $b_{k(i)}$.

First, we discuss in more detail the commitment mechanism. The description of the commitment phase states that measuring $\hat{C}_0$ corresponds to the commitment to value $0$, while measuring $\hat{C}_1$ corresponds to the commitment to value $1$. From the expression of measuring observables in terms of the states $|0\rangle$ and $|1\rangle$, and the states orthogonal to them, given by equation (\ref{commitment_obs}), we see that when a bit value $b_k$, defining the qubit's quantum state $|b_k \rangle$, ``coincides'' with the measuring observable $\hat{C}_{c} = \hat{C}_{b_k}$, i.e. $b_k = c$, then the corresponding measurement outcome $r_i$, for which $k=k(i)$, coincides with the bit value $b_k$, i.e. $r_i = b_k$. This way, we can interpret the measurement outcome $r_i$ as Alice's inference of Bob's bit value $b_{k(i)}$: if the bit value and the observable ``coincide'', the inference will be right; otherwise, it will be random~\footnote{{We call this {\em ambiguous inference}: when measuring observable $\hat{C}_0$ and obtaining result $0$, we are not sure that the qubit was in the state $|0\rangle$, while obtaining $1$ means that the qubit's state was for sure $|1\rangle$ (and analogously for $\hat{C}_1$). In other words, when measuring $\hat{C}_c$ {\em all} qubits prepared in the state $|c\rangle$ will be among those for which the result was $c$ -- with the drawback that some among them will be prepared in the state $|1\rangle$. If we are interested in what we call {\em unambiguous inference} -- when measuring $\hat{C}_c$ and obtaining $c$ we want to be sure the state was $|c\rangle$, then we simply have to relabel measuring observables.}}. If by $p_c(r|b)$, with $c,r,b \in \{0,1\}$, we denote a conditional probability that a result $r$ is obtained when measuring observable $\hat{C}_c$ on state $|b \rangle$, then the overall conditional probabilities are given by the following expressions,
\begin{itemize}
	\item Alice measures $\hat{C}_0$:
\begin{equation}
\begin{aligned}
		\label{p_0}
		p_0(0|0) & = 1 \\
		p_0(1|0) & = 0 \\
		p_0(0|1) & = \cos ^2 \theta \\
		p_0(1|1) & = \sin ^2 \theta\,.
\end{aligned}
\end{equation}
    \item Alice measures $\hat{C}_1$:
\begin{equation}
\begin{aligned}
		\label{p_1}
		p_1(0|0) & = \sin ^2 \theta \\
		p_1(1|0) & = \cos ^2 \theta \\
		p_1(0|1) & = 0 \\
		p_1(1|1) & = 1\,.
\end{aligned}
\end{equation}
\end{itemize}

Thus, the above conditional probabilities give the signature of the commitment: if Alice measures $\hat{C}_0$, the statistics of her measurement outcomes will be given by (\ref{p_0}); otherwise, it will be given by (\ref{p_1}). It also serves as the proof of her commitment, during the opening phase: only if the statistics of $\{r_i \}$, with respect to $\{b_{k(i)} \}$, is ``close enough'' to $p_c(r|b)$, did Alice commit to the bit value $c$. If $n(r|b)$ is the number of measurements performed by Alice on qubits received in the state $|b\rangle$, {\em with outcome} $r$, and $n(b)$ is the {\em total number} of qubits received in the state $|b\rangle$, then define $q(r|b) = \frac{n(r|b)}{n(b)}$. By ``close enough'' to, say, $p_0(r|b)$ ($c=0$ commitment), we now mean that the probability $P(q||p_0)$ that the measurement statistics $q(r|b)$ were produced by a random source $p_0(r|b)$ is bigger than a certain threshold value $\alpha>0$ (Alice did measure $\hat{C}_0$). This represents Bob's criterion to accept Alice's commitment to $c = 0$, and analogously for $c = 1$. Moreover, for the protocol to be viable, we require that the statistics $q(r|b)$ which pass the test $P(q||p_0) > \alpha$ of committing to value $c=0$ are unlikely to be obtained by measuring $\hat{C}_1$, i.e. we require that 
\begin{equation}
	\label{viable}
P(q||p_0) > \alpha \Rightarrow P(q||p_1) < \beta,
\end{equation}
(and analogously for statistics that pass the test of committing to $c=1$). The security parameters $\alpha$ and $\beta$ are to be set by Bob and the protocol designer, respectively, depending on the particular equipment used and the desired level of confidence. 

Let us now discuss the protocol's ({\em practical}) security and show that it is indeed both binding and concealing. The protocol must guarantee that at a certain moment of time $T_1$, Alice commits to a bit value such that Bob cannot learn her commitment until she reveals it at time $T_2$, and Alice cannot change her decision after $T_1$, in particular at $T_2$.

Alice makes the commitment by measuring one of the two observables, during a period of time $[\tau_1,\tau_n]$. The protocol requires that Alice announces the arrival times of the qubits. To do so, she either performs her measurements straight away (meaning that she commits to a certain value), or she performs a non-demolition measurement, thus detecting the presence of a photon without either destroying it nor affecting its polarization state. However, photonic non-demolition measurements are still in their infancy stage (see \cite{Munro_2005}, \cite{Johnson_2010}). In \cite{Johnson_2010}, for instance, one performs a non-demolition measurement in a controlled environment (photons are in a confined cavity). This type of measurement is thus suitable for computational rather than optical cryptographic purposes. Furthermore, even if Alice had access to a non-demolition measuring apparatus, she would need, in order to perform her measurements later in time, to have access to stable long-term quantum memories. Despite recent efforts, current technology allows only for very short-term noisy memories, usually implemented by fiber optic cable (see Fig. 5 in \cite{Ng_2012} for an example): adding extra fiber optic cable makes the measurements much more prone to errors, and increases the qubit losses exponentially in the length of the cable. Consequently, Alice performs her measurements as soon as she receives the qubits sent to her by Bob, turning the arrival times of each qubit into the measurement times $\tau_i$ as well. Therefore, $\tau_n - \tau_1 \leq t_N - t_1 << T_2 - T_1$, and for all practical purposes we can say that the commitment in fact occurred at time $T_1$. Note that in the ideal case when all qubits sent by Bob arrive to Alice and are detected by her, $n = N$, we have $\tau_i = t_{k(i)} + vl$, with $v$ being the speed of the qubits and $l$ the distance between Alice and Bob.

For these same reasons, the protocol is binding: Alice must perform her measurements as the qubits arrive, and thus she must make a commitment at time $T_1$, and not later. Otherwise, she could have kept the qubits in a quantum memory and perform her measurements -- commit to a value -- later in time. On the other hand, as a consequence of the laws of quantum mechanics, there exists no measurement which would provide Alice with the knowledge of the states prepared by Bob for {\em all} of the received qubits: she cannot both know $p_0(r_i|0)$ and $p_1(r_i|1)$ for every $i$.
Thus, after performing her measurements, Alice cannot pass both the test  (\ref{p_0}) of committing to the value $0$ and the test (\ref{p_1}) of committing to the value $1$. Note that it is essential that the commitment tests  (\ref{p_0}) and  (\ref{p_1}) contain both $p_0(r_i|0)$ and $p_1(r_i|1)$, {\em as well as} $p_0(r_i|1)$ and $p_1(r_i|0)$, otherwise the protocol would not be binding. Indeed, Alice must both be able to identify states corresponding to her commitment choice {\em and} to have a proper statistics on states not corresponding to her choice. Otherwise, she could trivially pass both tests even without any measurements, by simply setting $r_i = 0$ for all $i$'s, in case she wants to pass the test of committing to the value $0$, and $r_i = 1$ otherwise.

Regarding the second security requirement, that of concealment, it is obvious that Alice's measurements reveal no information about her measurement outcomes, and thus of her commitment, to a spatially distant Bob. Sending entangled states would obviously not help, due to non-signaling and causality: Bob cannot infer the choice of Alice's local measurement by measuring his part of the entangled pair, spatially distant from Alice.

%Finally, we discuss the requirement of announcing the qubit times of arrival/measurement, which increases the protocol's security. Namely, in addition to the lack of stable quantum memories, it is also difficult, with today's technology, to perform non-demolition measurements on single qubits. It is especially true for the case of optical realizations, when qubits are encoded into polarization states of photons: it is not possible to detect a photon without destroying it (more generally, altering its state). Thus, Alice must perform her measurements as soon as the photon arrives, even if she had access to quantum memory, because it's the only way for her to learn the times of arrival $\tau_i$.

\section{Optical noise}
\label{Optical_noise}

The above discussion of the commitment mechanism, based on quantum complementarity, was done for the ideal case of noiseless channels and perfect sources and measurements~\footnote{{The only effects of the environment considered were the fact that not all of the qubits sent will arrive to Alice, and that some of those that arrive will not be detected at all. This was done only qualitatively, so that the protocol could be properly defined (mentioning that $n \leq N$).}}. In particular, the expressions (\ref{p_0}) and (\ref{p_1}) for conditional probabilities are obtained under this assumption. In this and the following section, we will discuss the case of a noisy environment. Regarding a future implementation of the protocol, in which the qubit states are encoded into the polarization of single photons~\cite{Silva_2010, Muga_2011, Almeida_2011, Almeida_2012}, we will discuss the case of optical realizations of the protocol. Nevertheless, our theoretical approach could be easily applied, with suitable small modifications, to other types of physical realizations as well. First, we will consider optical noise, while in the next section we will discuss non-optical effects, such as imperfect single-photon sources, losses during the transmission and imperfect detectors (detector efficiency and dark counts).

Note that in this protocol, unlike the case of quantum cryptography, we are interested in more than just one quantity. Quantum Bit Error Rate (QBER), used to study the effects of noise in quantum cryptography (see equations ($31$)-($33$) in \cite{Gisin-review}, page $166$), is the ratio between the wrongly measured versus the total number of qubits received, in case we measure in the {\em same basis} in which the qubits were prepared. In our case, though, we are interested in the ratios, i.e. the (conditional) probabilities of both the case of measurement in the same basis, and the case of measurement in a basis different from that in which the qubits were prepared. In the ideal case, the conditional probabilities were given by the expressions (\ref{p_0}) and (\ref{p_1}). In the following, we will present the corresponding conditional probabilities for the cases of depolarizing channel, bit-flip and phase-flip channels, and arbitrary unitary evolution. At the end of this section, we will combine the four contributions into a single one.

\subsection{Depolarizing channel}

The depolarizing channel is a model of white noise: with probability $(1-p)$, the state of a system (in our case qubit) stays the same, while with probability $p$ it becomes totally mixed. Note that in this case, the probability to obtain the result corresponding to the initial state is higher than $(1-p)$: even if, after passing the channel, the state of the system turns out to be totally mixed (which happens with probability $p$), there is still non-zero probability to obtain the result corresponding to the initial state. In this case, the probability to obtain the ``wrong" (e.g. opposite) result, when measuring in the same basis in which the qubits were prepared is:
\begin{equation}
p_0(1|0) = p_1(0|1) = p/2.
\end{equation}
This is nothing but the optical part of the QBER, given by equation (34) from \cite{Gisin-review}: $\mbox{QBER}_{\text{opt}}$ = (1-V)/2. Using this formula, where $V$ is the ``visibility" parameter, we get that $V = (1-p)$, which is precisely the probability that the state will pass the channel intact –- hence the name ``visibility" (a synonym, in a sense, of ``transparency").

The depolarizing channel has no preferred axis of action, in the sense that its action is the same in each basis of the systems's Hilbert space. The noise is the same along each axis, and is the most dominant type of noise/errors that occur, as it is a model of white noise. The Kraus decomposition (or the so-called operator-sum representation; see \cite{Nielsen}, page $360$, Section $8.2.3$) of the (super-)operator representing the depolarizing channel is, for the case of qubit states, given by:
\begin{equation}
	\label{depolarizing}
	\mathcal{E}_d (\hat{\rho})= (1-p) \hat{\rho} + p \frac{\hat{I}}{2} = (1 - \frac{3}{4}) \hat{\rho} + \frac{p}{4} \sum_{i=1}^3 \hat{\sigma}_i \hat{\rho} \hat{\sigma}_i\,,
\end{equation}
where $\hat{\rho}$ is a general mixed state representing the initial qubit state, $\hat{I}$ is the identity operator, and $\hat{\sigma}_i$ are the standard Pauli operators. Note that the first equality is the general definition, while the second one is a particular expression for a two-dimensional qubit case. %The qubit expression is obtained using the Bloch representation of qubit states $\hat{\rho} = \frac{1}{2}(\hat{I} + \vec{r}\cdot\hat{\vec{\sigma}})$, where $\vec{r}$ is a $3$D Bloch vector with $|\vec{r}|\leq 1$, and the usual (anti)commutation rules for the algebra of Pauli matrices (see \cite{Nielsen}, page $378$,  Section $8.3.4$).

Using the above definition (\ref{depolarizing}), one obtains the relevant conditional probabilities, analogous to (\ref{p_0}) and (\ref{p_1}) obtained for the ideal case ($\eta = \mbox{QBER}$ represents the optical part of the quantum bit error rate):
\begin{itemize}
	\item Alice measures $\hat{C}_0$:
\begin{equation}
\begin{aligned}
		\label{p_0-depolarizing}
		p_0(0|0) & = 1 - \frac{p}{2} = 1 - \eta \\
		p_0(1|0) & = \frac{p}{2} = \eta \\
		p_0(0|1) & = (1-p) \cos ^2 \theta + \frac{p}{2} = (1 - 2\eta )\cos^2 \theta + \eta \\
		p_0(1|1) & = (1-p) \sin ^2 \theta + \frac{p}{2} = (1 - 2\eta )\sin^2 \theta + \eta\,.
\end{aligned}
\end{equation}
    \item Alice measures $\hat{C}_1$:
\begin{equation}
\begin{aligned}
		\label{p_1-depolarizing}
		p_1(0|0) & = (1-p) \sin ^2 \theta + \frac{p}{2} = (1 - 2\eta )\sin^2 \theta + \eta \\
		p_1(1|0) & = (1-p) \cos ^2 \theta + \frac{p}{2} = (1 - 2\eta )\cos^2 \theta + \eta \\
		p_1(0|1) & = \frac{p}{2} = \eta \\
		p_1(1|1) & = 1 - \frac{p}{2} = 1 - \eta\,.
\end{aligned}
\end{equation}
\end{itemize}

\subsection{Bit-flip channel}

The other two types of channels, bit- and phase-flip, are basis dependent. This means that, in the case of a bit-flip in the $\mathcal{B}_0$ basis, it flips (changes) the state $|0\rangle$ into $|0^\perp\rangle$ (and vice-versa) with probability $p$, where, by construction $\langle 0|0^\perp\rangle = 0$. But, if the state is a general superposition $a|0\rangle + b|0^\perp\rangle$, the flipped state, $b|0\rangle + a|0^\perp\rangle$, will not in general be orthogonal to the initial state  $a|0\rangle + b|0^\perp\rangle$, so it will not be a bit-flip in other bases. Therefore, such noise/errors are expected to occur in cases where we can isolate a preferable axis (and thus a basis), which is the case of a measurement of an observable (or a preparation of a certain state, which is a basis state of a certain observable). The operator-sum representation of the bit-flip channel is:
\begin{equation}
	\label{E_p}
	\mathcal{E}_{b_0} (\hat{\rho})= (1-p) \hat{\rho} + p \hat{\sigma}_{x_0} \hat{\rho} \hat{\sigma}_{x_0}.
\end{equation}

For simplicity we will not present the relevant conditional probabilities here, since they can be easily derived from the general formula \eqref{p_0-tot} for combined noises, given in Subsection \ref{total_noise_section}.

%The relevant conditional probabilities are:
% \begin{itemize}
% 	\item Alice measures $\hat{C}_0$:
% \begin{equation}
% \begin{aligned}
% 		\label{p_0-bit-flip}
% 		p_0(0|0) & = 1 - p \\
% 		p_0(1|0) & = p \\
% 		p_0(0|1) & = \cos ^2 \theta - p\cos 2\theta \\
% 		p_0(1|1) & = \sin ^2 \theta + p\cos 2\theta\,.
% \end{aligned}
% \end{equation}
%     \item Alice measures $\hat{C}_1$:
% \begin{equation}
% \begin{aligned}
% 		\label{p_1-bit-flip}
% 		p_1(0|0) & = \sin ^2 \theta + p\cos 2\theta \\
% 		p_1(1|0) & = \cos ^2 \theta - p\cos 2\theta \\
% 		p_1(0|1) & = p - p(\sin 2\theta)^2 (\cos\phi)^2 \\
% 		p_1(1|1) & = (1-p) + p(\sin 2\theta)^2 (\cos\phi)^2\,.
% \end{aligned}
% \end{equation}
% \end{itemize}

Note that the above channel flips the basis states $|0\rangle$ into $|0^\perp\rangle$ (and vice-versa), such that the matrix representation of the operator $\hat{\sigma}_{x_0}$ in the basis $\mathcal{B}_0 = \{|0\rangle,|0^\bot\rangle \}$ is the Pauli matrix $\sigma_x = [\hat{\sigma}_{x_0}]_{\mathcal{B}_0}=\left[ \begin{array}{cc} 0 & 1 \\ 1 & 0 \end{array}\right]$. We could also consider a bit-flip channel where states $|1\rangle$ and $|1^\perp\rangle$ are flipped, using $\hat{\sigma}_{x_1} = |1\rangle\langle 1^\perp|  + |1^\perp\rangle \langle 1|$ instead of $\hat{\sigma}_{x_0}$. %Such channel $\mathcal{E}_{b_1}$ would be given by the formula analogous to (\ref{E_p}), with $\hat{\sigma}_{x_1} = |1\rangle\langle 1^\perp|  + |1^\perp\rangle \langle 1|$ instead of $\hat{\sigma}_{x_0}$, and the expressions for the corresponding conditional probabilities would be different.

\subsection{Phase-flip channel}

The second basis-dependent operation to model the noise is the phase-flip channel. It ``flips" the phase of one of the two basis vectors. Below, as in the previous case, we fix the basis $\mathcal{B}_0$ and represent the flip of the phase of $|0^\bot\rangle$ by the operator $\hat{\sigma}_{z_0}$ whose matrix representation is again a Pauli matrix $\sigma_z = [\hat{\sigma}_{z_0}]_{\mathcal{B}_0}=\left[ \begin{array}{cc} 1 & 0 \\ 0 & -1 \end{array}\right]$.
The operator-sum representation of the phase-flip channel is:
\begin{equation}
	\mathcal{E}_{p_0} (\hat{\rho})= (1-p) \hat{\rho} + p \hat{\sigma}_{z_0} \hat{\rho} \hat{\sigma}_{z_0}.
\end{equation}

% The relevant conditional probabilities are:
% \begin{itemize}
% 	\item Alice measures $\hat{C}_0$:
% \begin{equation}
% \begin{aligned}
% 		\label{p_0-phase-flip}
% 		p_0(0|0) & = 1 \\
% 		p_0(1|0) & = 0 \\
% 		p_0(0|1) & = \cos ^2 \theta \\
% 		p_0(1|1) & = \sin ^2 \theta\,.
% \end{aligned}
% \end{equation}
%     \item Alice measures $\hat{C}_1$:
% \begin{equation}
% \begin{aligned}
% 		\label{p_1-phase-flip}
% 		p_1(0|0) & = \sin ^2 \theta \\
% 		p_1(1|0) & = \cos ^2 \theta \\
% 		p_1(0|1) & = p(\sin 2\theta)^2 \\
% 		p_1(1|1) & = 1-p(\sin 2\theta)^2\,.
% \end{aligned}
% \end{equation}
% \end{itemize}

Again, the relevant conditional probabilities are shown in \eqref{p_0-tot}. In analogy with the bit-flip channel, here as well we could consider a channel  $\mathcal{E}_{p_1}$ given by the operator $\hat{\sigma}_{z_1} = |1\rangle\langle 1|  - |1^\perp\rangle \langle 1^\perp|$. %Note that a phase-flip, occurring in base $\mathcal{B}_0$ or $\mathcal{B}_1$, affects the conditional probabilities only when both the eigenbasis of the measurement observable and the state sent by Bob are different from the basis of the phase flip: in the above example of the phase flip $\mathcal{E}_{p_0}$ occurring in the $\mathcal{B}_0$ basis, the only conditional probabilities affected by the phase-flip are $p_1(\ast|1)$, when the measurement is performed in the $\mathcal{B}_1$ basis, on the state $|1 \rangle$.

\subsection{Unitary evolution}

Finally, the unitary evolution could be used to model the cases for which we have a constant ``rotation" of the state of a system. For example, we may send qubits as photons through an optical fibre which, due to bad twisting, rotates the polarization angle by a fixed ratio per unit length~\cite{Muga_2006}. The unitary evolution is given by:
\begin{equation}
	\mathcal{E}_u (\hat{\rho})= \hat{U} \hat{\rho} \hat{U}^\dag\,,
\end{equation}
where the arbitrary $U(1)$ unitary operator is, up to an irrelevant global phase, given by its matrix representation (say, in the $\mathcal{B}_0$ basis):
\begin{equation}
	[\hat{U}]_{\mathcal{B}_0} = \left[ \begin{array}{cc} e^{i\lambda}\cos \alpha & -e^{-i\mu}\sin \alpha \\ e^{i\mu}\sin \alpha & e^{-i\lambda}\cos \alpha \end{array}\right]
\end{equation}
with $\alpha \in [0, \pi /2]$ and $\lambda, \mu \in [0, 2\pi)$~\cite{U2}. Note that $\sin^2 \alpha  = \eta$ is the QBER {\em in the $\mathcal{B}_0$ basis.} 

The relevant conditional probabilities are given in \eqref{p_0-tot}.

% The relevant conditional probabilities are:
% \begin{itemize}
% \item Alice measures $\hat{C}_0$:
% \begin{equation}
% \begin{aligned}
% 		\label{p_0-unitary}
% 		p_0(0|0) & = \cos^2 \alpha = 1 - \eta \\
% 		p_0(1|0) & = \sin^2 \alpha = \eta \\
% 		p_0(0|1) & = (1-\eta) \cos ^2 \theta + \eta\sin^2\theta - \\ &\sqrt{\eta(1-\eta)}\sin 2\theta \cos (\phi - \lambda - \mu) \\
% 		p_0(1|1) & = (1-\eta) \sin ^2 \theta + \eta\cos^2\theta + \\ &\sqrt{\eta(1-\eta)}\sin 2\theta \cos (\phi - \lambda - \mu)\,.
% \end{aligned}
% \end{equation}
% \item Alice measures $\hat{C}_1$: the conditional probabilities expressed in terms of parameters $\alpha$, $\lambda$ and $\mu$ are rather lengthy, in particular the two $p_1(\ast |1)$ probabilities. Nevertheless, expressed in terms of $\alpha^\prime$, $\lambda^\prime$ and $\mu^\prime$ (and consequently $\eta^\prime$), given by the matrix elements of $\hat{U}$ in the $\mathcal{B}_1$ basis, the expressions for $p_1(\ast | \ast )$ can be obtained from $p_0(\ast | \ast)$, by exchanging $0$ and $1$.
% \end{itemize}

\subsection{Total optical noise accumulated during the emission, transmission and measurement}
\label{total_noise_section}

The next, and final, step in modeling the optical part of the noise is combining the above four contributions in a single set of formulas for conditional probabilities. Our approach is the following. The whole apparatus consists of three parts: the Sender (Bob) performing the state preparation using the source of photons, the transmission environment (transmission through the space, optical fibre, etc.), and the receiver (Alice) performing the measurement using essentially detectors and beam-splitters.

In each of the three parts, we can have some of the above described types of noise. The white noise occurs with particular probabilities $p^p_d$, $p^t_d$ and $p^m_d$ ($p$, $t$ and $m$ stand for preparation, transmission and measurement, and $d$ for depolarizing channel), characteristic to the equipment and the environment.
%In addition to the white noise, we can also have the bit- and the phase-flip channels, given by their corresponding probabilities (three for each channel, as in the case of the white noise). Note that, for simplicity, we omit the additional label referring to the type of noise. Finally, a certain (usually systematic) unitary rotation can in general happen during each of the three phases.

The white noise is a generic type of noise that affects all types of instruments and environments. During the preparation and the measurement, we can also have basis-dependent noises: when preparing the $|0\rangle$ state, we can have a bit-flip and a phase-flip in the basis $\mathcal{B}_0 = \{|0\rangle,|0^\bot\rangle \}$, characteristic for this particular preparation procedure. They occur with the corresponding probabilities $p_b$ and $p_p$ ($b$ and $p$ standing for the bit and phase, respectively; the upper labels are omitted for simplicity). When preparing the $|1\rangle$ state, bit- and phase-flips occur in the basis $\mathcal{B}_1 = \{|1\rangle,|1^\bot\rangle \}$, with the {\em same} probabilities $p_b$ and $p_p$ as in the previous case, since the two-state preparation apparatuses are rotated with respect to each other (assuming spatial isotropy). In general, bit- and phase-flips could occur along a general axis during the transmission as well, but this is a highly unlikely scenario as usually the transmission environment (space, optical fibre) has no preferential axes (bases) and thus the noise is not likely to be biased in this manner. Finally, a unitary evolution could occur during the transmission, while it is unlikely to happen in sources and detectors, and is thus ignored in the preparation and measurement phases.

Therefore, the overall state of a qubit after ``passing through" the preparation apparatus, transmission environment, and the measurement apparatus, just before the detection, can be obtained by the consecutive application of the following channels:
\begin{itemize}
	\item depolarizing, bit-flip and phase-flip channels, each with different probabilities $(p^p_d,p^p_b,p^p_p)$, in the preparation apparatus,
    \item depolarizing channel, with probability $p^t_d,$ and unitary rotation,  during the transmission, and
    \item depolarizing, bit-flip and phase-flip channels, each with different probabilities $(p^m_d,p^m_b,p^m_p)$, during the measurements.
\end{itemize}

In each part of the apparatus (sender, transmission environment and receiver) different channels model different noises that occur at the same time, which is assured by their commutativity. Indeed, all the commutation relations needed are satisfied: depolarizing, bit- and phase-flips channels commute with each other (a consequence of commutation relations for the Pauli matrices and the particular Kraus representations of the three channels in terms of Pauli matrices, see \cite{Nielsen}, for example). Thus, the order of their application during the preparation and measurement are irrelevant. In particular, we can apply the depolarizing channel occurring during the preparation just before the transmission, and the one occurring during the measurement just after the transmission. Depolarizing channel and unitary evolution commute as well (the white noise is isotropic), so that we can treat the white noise by only one parameter: since $\mathcal{E}_d = \mathcal{E}_d^m \circ \mathcal{E}_d^t \circ \mathcal{E}_d^p$, we have $p_d = p_d^m + p_d^t + p_d^p$. On the other hand, the bit- and phase-flips {\em do not} commute with the unitary evolution (as no axis-dependent operation commutes with the unitary evolution, in general).

Combining the overall noise in the apparatus, depending on the preparation procedure (preparing either $|0\rangle$ of the $\mathcal{B}_0$ basis, or $|1\rangle$ of the $\mathcal{B}_1$ basis), we have four distinct channels:
\begin{itemize}
	\item $\mathcal{E}_{00} = \mathcal{E}_{b_0}^{m} \circ \mathcal{E}_{p_0}^{m} \circ \mathcal{E}_d \circ \mathcal{E}_u^{t} \circ \mathcal{E}_{b_0}^{p} \circ \mathcal{E}_{p_0}^{p}$, when measuring $\hat{C}_0$ and preparing $|0\rangle$,
    \item $\mathcal{E}_{01} = \mathcal{E}_{b_0}^{m} \circ \mathcal{E}_{p_0}^{m} \circ \mathcal{E}_d \circ \mathcal{E}_u^{t} \circ \mathcal{E}_{b_1}^{p} \circ \mathcal{E}_{p_1}^{p}$, when measuring $\hat{C}_0$ and preparing $|1\rangle$,
    \item $\mathcal{E}_{10} = \mathcal{E}_{b_1}^{m} \circ \mathcal{E}_{p_1}^{m} \circ \mathcal{E}_d \circ \mathcal{E}_u^{t} \circ \mathcal{E}_{b_0}^{p} \circ \mathcal{E}_{p_0}^{p}$, when measuring $\hat{C}_1$ and preparing $|0\rangle$,
	\item $\mathcal{E}_{11} = \mathcal{E}_{b_1}^{m} \circ \mathcal{E}_{p_1}^{m} \circ \mathcal{E}_d \circ \mathcal{E}_u^{t} \circ \mathcal{E}_{b_1}^{p} \circ \mathcal{E}_{p_1}^{p}$, when measuring $\hat{C}_1$ and preparing $|1\rangle$.
\end{itemize}
%Note that in the above case of preparing only the states $|0\rangle$ and $|1\rangle$, and measuring either $\hat{C}_0$ or $\hat{C}_1$, the possible bit-flips do not affect the conditional probabilities: the phase-flips during the preparation do not affect the prepared state (rather the one orthogonal to it), while the flips of the phase, occurring in the same basis as the eigenbasis of the measurement observable, change only the sign of the off-diagonal elements of the qubit's density matrix, thus preserving the probabilities (the diagonal elements).

The calculation of the corresponding conditional probabilities is rather lengthy, but quite straightforward. We present the final expressions for the conditional probabilities when Alice measures $\hat{C}_0$:
\begin{widetext}
\begin{equation}
\begin{aligned}
\label{p_0-tot}
     p_0(0|0) & = \frac{1}{2} \bigg[1 + (1- p_d) (1 - 2 p_b^m) (1 - 2 p_b^p) \cos 2 \alpha \bigg]\,, \\
     p_0(1|0) & = \frac{1}{2} \bigg[1 - (1- p_d) (1 - 2 p_b^m) (1 - 2 p_b^p) \cos 2 \alpha \bigg]\,. \\
     p_0(0|1) & = \frac{1}{2} [1 + (1- p_d) (1 - 2 p_b^m) (1 - 2 p_b^p) \cos 2 \alpha \cos 2 \theta - (1- p_d) (1 - 2 p_b^m) (1 - 2 p_b^p) \sin 2 \alpha \sin 2 \theta \cos(\phi - \lambda - \mu)]\,,\\
     p_0(1|1) & = \frac{1}{2} \bigg[1 - (1- p_d) (1 - 2 p_b^m) (1 - 2 p_b^p) \cos 2 \alpha \cos 2 \theta + (1- p_d) (1 - 2 p_b^m) (1 - 2 p_b^p) \sin 2 \alpha \sin 2 \theta \cos(\phi - \lambda - \mu)\bigg]\,.
\end{aligned}
\end{equation}
\end{widetext}
% \begin{widetext}
% \begin{eqnarray}
% \label{p_0-tot}
%      p_0(0|0) & = & \frac{1}{2} \bigg[1 + (1- p_d) (1 - 2 p_b^m) (1 - 2 p_b^p) \cos 2 \alpha \bigg] \nonumber \\
%      p_0(1|0) & = & \frac{1}{2} \bigg[1 - (1- p_d) (1 - 2 p_b^m) (1 - 2 p_b^p) \cos 2 \alpha \bigg] \\
%      p_0(0|1) & = & \frac{1}{2} [1 + (1- p_d) (1 - 2 p_b^m) (1 - 2 p_b^p) \cos 2 \alpha \cos 2 \theta - (1- p_d) (1 - 2 p_b^m) (1 - 2 p_b^p) \sin 2 \alpha \sin 2 \theta \cos(\phi - \lambda - \mu)] \nonumber\\
%      p_0(1|1) & = & \frac{1}{2} \bigg[1 - (1- p_d) (1 - 2 p_b^m) (1 - 2 p_b^p) \cos 2 \alpha \cos 2 \theta + (1- p_d) (1 - 2 p_b^m) (1 - 2 p_b^p) \sin 2 \alpha \sin 2 \theta \cos(\phi - \lambda - \mu)\bigg]. \nonumber
% \end{eqnarray}
% \end{widetext}
The results for the case when Alice measures $\hat{C}_1$ can be obtained from the above ones by exchanging labels $0$s with $1$s in the conditional probabilities, for example $p_0(1|0) = p_1(0|1)$, etc. Note that the depolarizing and bit-flip coefficients occur in the same way in all expressions, as $(1- p_d) (1 - 2 p_b^m) (1 - 2 p_b^p)$. Moreover, the effects of both bit-flips (during the state preparation and during the measurement) have the same form as that of a depolarizing channel. Indeed, by introducing $b = 1 - (1 - 2 p_b^m) (1 - 2 p_b^p)$, we obtain the joint depolarizing -- bit-flip coefficient in a symmetric form $(1-p_d)(1-b)$. Note that, since $p_b^{p/m} \in [0,1/2]$, we have $b\in [0,1]$; the ranges of the two coefficients coincide.

\subsection{Distinguishability between conditional probabilities corresponding to different commitment choices}

Each discrete probability distribution $p(i)$, with $i = 1, \ldots n$, can be seen as a vector $p = (p_1, p_2, \ldots p_n)$, whose coordinates $p_i$ are the probabilities, $p_i = p(i)$. Yet, there is a more suitable representation as a vector $p = (p_1, p_2, \ldots p_n)$, where the coordinates $p_i$ are square roots of the probabilities, $p_i = \sqrt{p(i)}$. The motivation for this representation is the following: with the standard scalar product, $p \cdot q = (p,q) = \langle p | q \rangle = \sum_i p_i q_i = \sum_i \sqrt{p(i)q(i)}$, all vectors representing probability distributions have unit norm, due to the normalization of probabilities to one. On the other hand, a way to quantify distinguishability between two probability distributions $p$ and $q$ is given by the {\em fidelity} $F(p,q) \equiv \sum_i \sqrt{p(i)q(i)}$ (known also as the {\em Bhattacharyya coefficient} \cite{fidelity}), which is nothing but the scalar  product we just introduced, $p \cdot q = \sum_i \sqrt{p(i)q(i)} = F(p,q)$.
The more similar the two probabilities are, the bigger the scalar product is ($1$ when they are identical); the more different, i.e. distinguishable, they are, the smaller the scalar product is (the most distingushable being the orthogonal ones).

We can use this measure of the probability distinguishability to study the influence of noise to the protocol's performance. In this and the following sections, we will not consider the effects of a possible unitary rotation during the transmission, as it represents a systematic error that can be compensated. Nevertheless, we hope the above results on the conditional probabilities with the influence of a possible unitary rotation could be useful in detecting and eliminating such a systematic error.

When measuring $\hat{C}_0$, we get two probability distributions, each conditioned by the input state $|0\rangle$ or $|1\rangle$: one is $p_0(\ast|0) = (\sqrt{p_0(0|0)}, \sqrt{p_0(1|0)})$, the other $p_0(\ast|1) = (\sqrt{p_0(0|1)}, \sqrt{p_0(1|1)})$.

When measuring $\hat{C}_1$, we get other two probability distributions, again each conditioned by the input state $|0\rangle$ or $|1\rangle$: one is $p_1(\ast|0) = (\sqrt{p_1(0|0)}, \sqrt{p_1(1|0)})$, the other $p_1(\ast|1) = (\sqrt{p_1(0|1)}, \sqrt{p_1(1|1)})$.

The stronger the noise is, the more the resulting conditional probabilities diverge form the ideal case, given by (\ref{p_0}) and (\ref{p_1}), approaching to a pair of totally balanced conditional probabilities. Thus, we may consider the average fidelity between the corresponding probability distributions for the noiseless case and the case of a noise given by the channel $\mathcal{E}$. If $p_c(r|b)$ and $p_c^\mathcal{E}(r|b)$ are the probabilities for the ideal noiseless case and the case with a noise given by the channel $\mathcal{E}$, respectively ($c,r,b \in \{ 0,1 \}$), then the average fidelity between the four probability distributions is
\begin{equation}
\begin{aligned}
\label{fid_noise}
 \langle F(\mathcal{E}) \rangle &= \frac{1}{4} \left[F\left(\mathcal{E}; \hat{C}_0, |0\rangle \right) +  F\left(\mathcal{E}; \hat{C}_0, |1\rangle \right) + \right.\\
 &\qquad \left. {} F\left(\mathcal{E}; \hat{C}_1, |0\rangle \right) +  F\left(\mathcal{E}; \hat{C}_1, |1\rangle \right)\right]\,,
\end{aligned}
\end{equation}
where the four fidelities between the four pairs of probability distributions, each obtained for the case of Alice measuring $\hat{C}_c$, when the state sent by Bob was $|b\rangle$ are:
\begin{widetext}
\begin{equation}
\begin{aligned}
	\label{fid_noise_4}
 F \left(\mathcal{E}; \hat{C}_0, |0\rangle \right) & = & F \left( p_0(\ast |0), p_0^\mathcal{E}(\ast |0) \right) = \left( \sqrt{p_0(0|0)p_0^\mathcal{E}(0|0)} + \sqrt{p_0(1|0)p_0^\mathcal{E}(1|0)} \right)\,, \\
 F \left(\mathcal{E}; \hat{C}_0, |1\rangle \right) & = & F \left( p_0(\ast |1), p_0^\mathcal{E}(\ast |1) \right) = \left( \sqrt{p_0(0|1)p_0^\mathcal{E}(0|1)} + \sqrt{p_0(1|1)p_0^\mathcal{E}(1|1)} \right)\,, \\
 F \left(\mathcal{E}; \hat{C}_1, |0\rangle \right) & = & F \left( p_1(\ast |0), p_1^\mathcal{E}(\ast |0) \right) = \left( \sqrt{p_1(0|0)p_1^\mathcal{E}(0|0)} + \sqrt{p_1(1|0)p_1^\mathcal{E}(1|0)} \right)\,, \\
 F \left(\mathcal{E}; \hat{C}_1, |1\rangle \right) & = & F \left( p_1(\ast |1), p_1^\mathcal{E}(\ast |1) \right) = \left( \sqrt{p_1(0|1)p_1^\mathcal{E}(0|1)} + \sqrt{p_1(1|1)p_1^\mathcal{E}(1|1)} \right)\,.
\end{aligned}
\end{equation}
\end{widetext}
The bigger the above expected fidelity is (the more similar the actual probability distributions are to the ideal noiseless ones), the higher is the protocol's security.

One could also analyze the intrinsic properties of the conditional probabilities $p_c^\mathcal{E}(\ast |b)$, obtained when a noise $\mathcal{E}$ is present (for simplicity, when considering the intrinsic properties of distributions in noisy environments, we drop the superscript $\mathcal{E}$). As mentioned before, when presenting the protocol, each choice of Alice's measurement can serve to infer the qubit state, prepared by Bob. Thus, whatever the observable $\hat{C}_c$ she measures, the corresponding distributions obtained for the case when the prepared state is $|0\rangle$, and when it is $|1\rangle$, should be as distinguishable as possible. The fidelities between these two pairs of probability distributions are:
\begin{equation}
\begin{aligned}
F(p_0(\ast|0), p_0(\ast|1)) & = \sqrt{p_0(0|0) p_0(0|1)} + \sqrt{p_0(1|0) p_0(1|1)}\,, \\
F(p_1(\ast|0), p_1(\ast|1)) & = \sqrt{p_1(0|0) p_0(0|1)} + \sqrt{p_1(1|0) p_0(1|1)}\,.
\end{aligned}
\end{equation}
The average fidelity between the probability distributions obtained when sending the state $|0\rangle$ and the state $|1\rangle$ is then:
\begin{equation}
\begin{aligned}
\label{fid_states}
 \langle F(|0\rangle, |1\rangle) \rangle &= 1/2 \left[F\left(p_0(\ast|0), p_0(\ast|1) \right) + \right.\\
 &\qquad \left. {} F( p_1(\ast|0), p_1(\ast|1))\right]\,.
\end{aligned}
\end{equation}
The smaller this average fidelity is, the more distinguishable the two distributions are, thus the better can Alice infer which state was sent by Bob, and the protocol security is better.

Finally, note that Alice's choice of measurement must produce two rather different conditional probability distributions $p_0(\ast |b)$ and $p_1(\ast |b)$. Only then can her commitment be imprinted in the set of her measurement outcomes, so that Bob can learn Alice's commitment during the opening phase, and Alice cannot change decision (the protocol is binding). The average fidelity between  the two sets of probability distributions, obtained when measuring $\hat{C}_0$, and $\hat{C}_1$, respectively, is:
\begin{equation}
\begin{aligned}
\label{fid_observables}
 \langle F(\hat{C}_0, \hat{C}_1)\rangle &= 1/2 \left[F\left(p_0(\ast|0), p_1(\ast|0)\right) + \right.\\
 &\qquad \left. {} F(p_0(\ast|1), p_1(\ast|1))\right]\,,
\end{aligned}
\end{equation}
where:
\begin{equation}
\begin{aligned}
F( p_0(\ast |0), p_1(\ast |0) ) & = \sqrt{ p_0(0|0) p_1(0|0) } + \sqrt{ p_0(1|0) p_1(1|0) }\,, \\
F( p_0(\ast |1), p_1(\ast |1) ) & = \sqrt{ p_0(0|1) p_1(0|1) } + \sqrt{ p_0(1|1) p_1(1|1) }\,.
\end{aligned}
\end{equation}
Again, the smaller this expected fidelity is, the more distinguishable the two distributions are, which results in higher security of the protocol: the more distingushable are the actions of Alice, the more secure is her commitment (the choice of her action).

Note that in the noiseless case, we have that $\langle F ( |0\rangle, |1\rangle) \rangle = \cos^2 \theta$, while $\langle F ( \hat{C}_0, \hat{C}_1 ) \rangle = \sin^2 \theta$. Thus, according to the first criterion, the best state distinguishability is, as expected, achieved for $\theta = \pi /2$, while the highest measurement distinguishability is achieved for $\theta = 0$ (again, this is a rather trivial fact when Bob sends qubits in only one state, which corresponds to result $0$ when measuring $\hat{C}_0$ and $1$ when measuring $\hat{C}_1$ -- the two observables represent the same physical property, with its outcomes being re-labelled). The two opposed security requirements become equal for $\theta = \pi /4$, which matches the optimal value for the angle between the states sent by Bob.

\begin{figure*}[t!]
	\centering
    \includegraphics[width=0.45\textwidth]{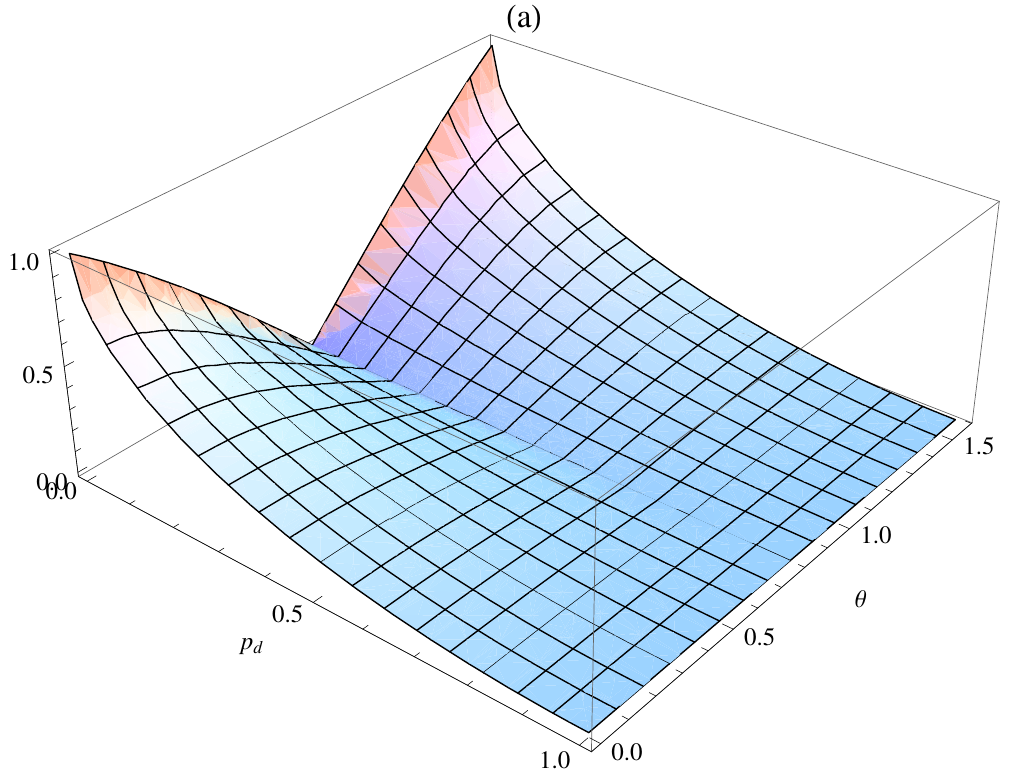}\quad
    \includegraphics[width=0.45\textwidth]{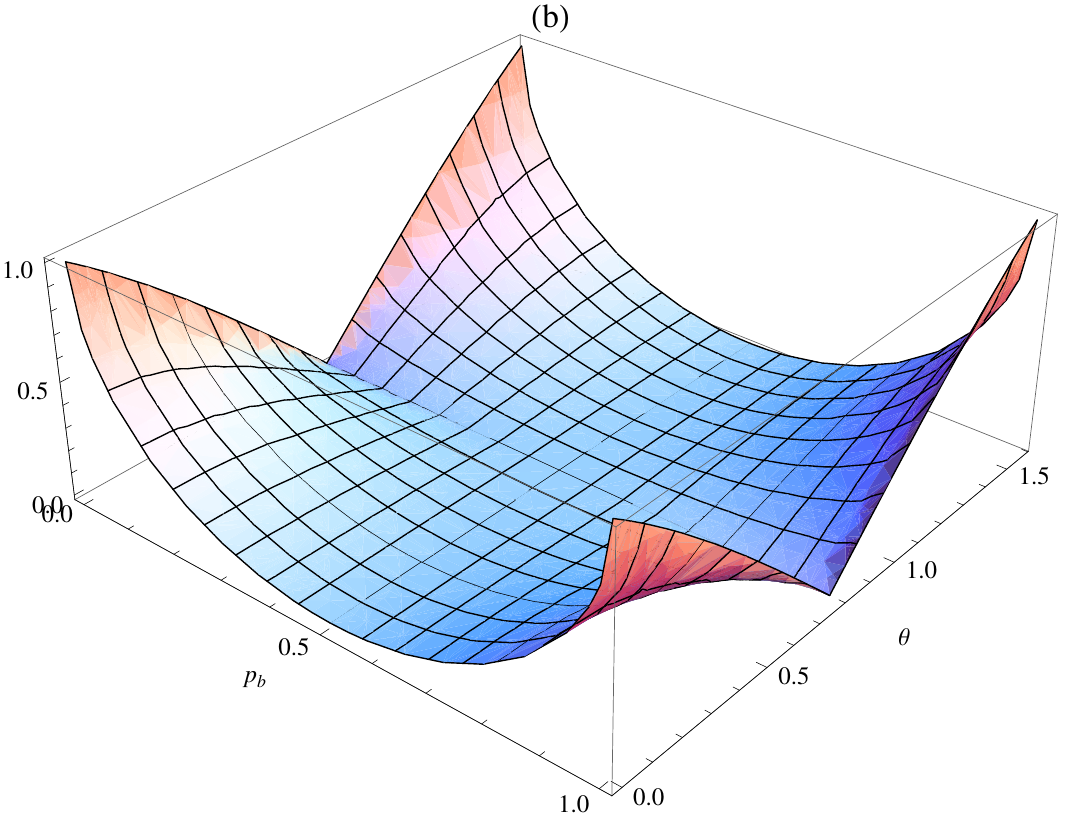}
	\caption{\label{DC_channel_optimal_l}(Color online) (a) Graph of $| \langle F ( |0\rangle, |1\rangle) \rangle - \langle F ( \hat{C}_0, \hat{C}_1 ) \rangle |$, as a function of $\theta$ and $p_d$. (b) Graph of $| \langle F ( |0\rangle, |1\rangle) \rangle - \langle F ( \hat{C}_0, \hat{C}_1 ) \rangle |$, as a function of $\theta$ and $p_b$.}
\end{figure*}

The same happens for noisy channels. In particular, in Fig.~\ref{DC_channel_optimal_l}(a) is plotted the graph of $| \langle F ( |0\rangle, |1\rangle) \rangle -\langle F ( \hat{C}_0, \hat{C}_1 ) \rangle |$, as a function of $\theta$ and $p_d$ in the case of a depolarizing channel. As shown, the optimal choice of $\theta$ is $\theta =\pi /4$, unless $p_d =1$, in which case the measurement results are completely random, and consequently any $\theta$ will yield the same behavior. An analogous phenomenon occurs in the bit-flip and the phase-flip channels (see Fig.~\ref{DC_channel_optimal_l}), in which cases complete randomness is achieved by setting $p_b =1/2$ and $p_p=1/2$, respectively (note that on this plot we extended the domain of the bit-flip coefficient to $[0,1]$ obtaining the plot symmetric around the value $p_b =1/2$).

\begin{figure*}[t!]
	\centering
    \includegraphics[width=0.45\textwidth]{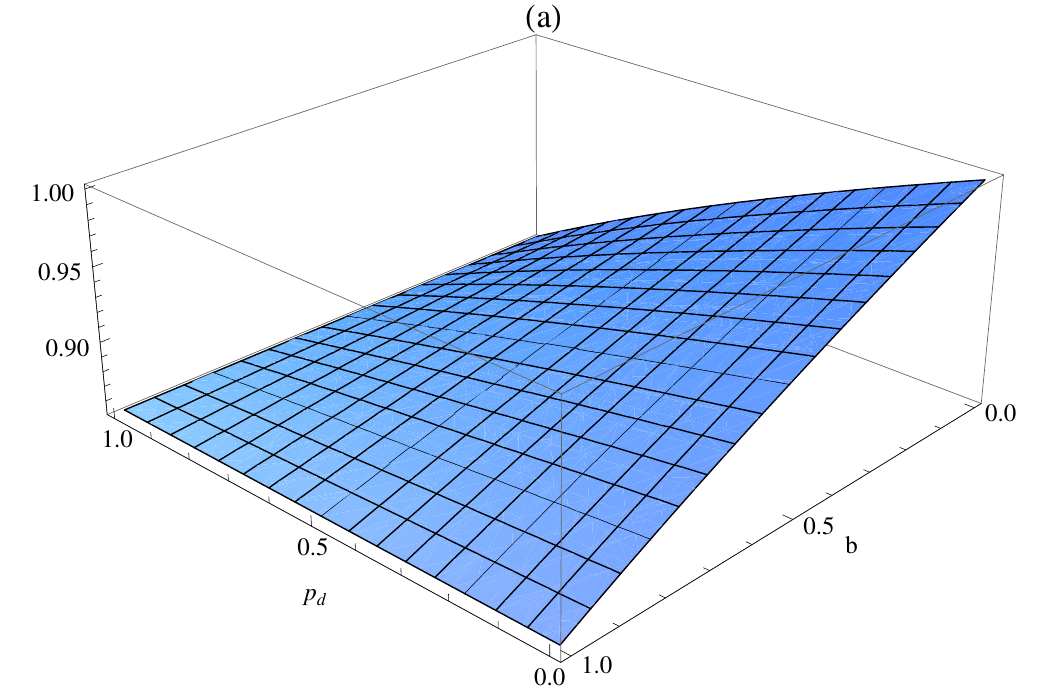}\quad
    \includegraphics[width=0.45\textwidth]{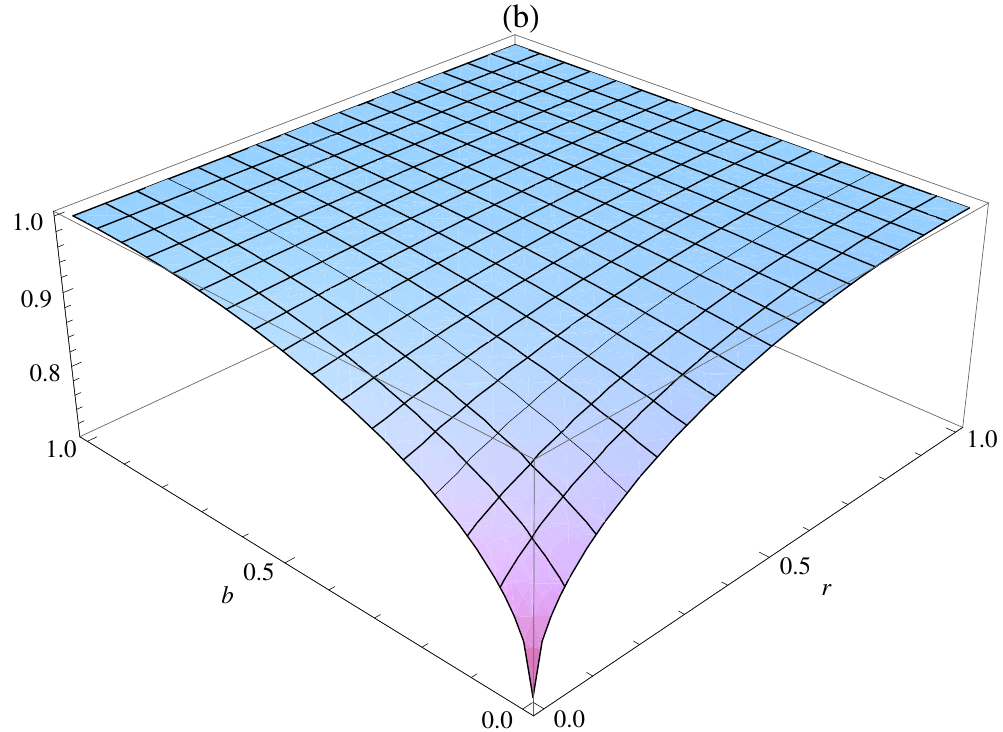}
	\caption{\label{Fid_noisy_noiseless_l}(Color online) Fidelity $\langle F(\mathcal{E}) \rangle$ between noisy and noiseless channels (a), and Fidelity $\langle F ( |0\rangle, |1\rangle) \rangle =\langle F ( \hat{C}_0, \hat{C}_1 ) \rangle$ within the noisy channel (b), as functions of the depolarizing coefficient $p_d$ and the joint (preparation- and measurement-induced) bit-flip coefficient $b =1-(1-2p_b^p)(1-2p_b^m)$.}
\end{figure*}

Next we present the effects of noise on the average fidelity $\langle F(\mathcal{E}) \rangle$ between noisy and noiseless channels \eqref{fid_noise} - see Fig.~\ref{Fid_noisy_noiseless_l}(a) - and within the noisy channel itself,  $\langle F ( |0\rangle, |1\rangle) \rangle$ and $\langle F ( \hat{C}_0, \hat{C}_1 ) \rangle$, given by \eqref{fid_states} and \eqref{fid_observables}, respectively, for the optimal choice of $\theta = \pi/4$ (recall that in this case the two are equal) - see Fig.~\ref{Fid_noisy_noiseless_l}(b). In the plots, as noted in the previous section, the coefficient $b$ stands for $1-(1-2p_b^p)(1-2p_b^m)$, and represents the joint effect of the bit-flip channels in the preparation and measurement apparatuses. Both figures show the expected behavior. Adding noise gradually takes the probability distributions from a noiseless case to a completely random case. Also note that the effect of the bit-flip channel is the same as the effect of the depolarizing channel.

The other quantity widely used to measure how different probability distributions are is the {\em relative entropy} (also known as {\em Kullback-Leibler divergence} \cite{rel_ent}; for a review of the use of relative entropy in the field of quantum information, see \cite{vlatko_rel_ent}). For two probability distributions $\{ p_i \}$ and $\{ q_i \}$, the relative entropy between the two is given by:
\begin{equation}
S(p||q) = \sum_i p_i \ln \frac{p_i}{q_i}\,.
\end{equation}
%Here, $\ln$ represents the natural logarithm for base $e$. In case we want to quantify the relative entropy ``in bits'', i.e. use the logarithm for the base $2$, the above value should be rescaled by dividing it by $\ln 2 \approx 0.69$.
Although not formally a distance -- it is not symmetric with respect to its arguments -- it still can serve as a measure of distinguishability~\footnote{{Indeed, its infinitesimal form is a true metric, the so-called Fisher information metric (see for example~\cite{fisher}, page $87$).}}.
It determines the probability that a random source that emits symbols according to a probability distribution $\{ q_i \}$ will produce a sequence of symbols consistent with a source emitting according to $\{ p_i \}$ (see {\em Theorem} $4$ from~\cite{vlatko_rel_ent}).

Here as well we can consider the quantities analogous to those considered in the case of the fidelity, $\langle S(\mathcal{E}) \rangle$, $\langle S ( |0\rangle || |1\rangle) \rangle$ and $\langle S ( \hat{C}_0 || \hat{C}_1 ) \rangle$, given by expressions analogous to \eqref{fid_noise}, \eqref{fid_states} and \eqref{fid_observables}. Note that, since relative entropy is not symmetric, we can consider six rather than just three quantities. In the case of $\langle S(\mathcal{E}) \rangle$ though, only one of the two options is relevant to our study: that which quantifies the probability that the noisy environment and imperfect apparatus will reproduce results as in the ideal case given by \eqref{p_0} and \eqref{p_1}. We also note that in the noiseless case the state and measurement distinguishabilities, according to relative entropy, become equal for the optimal value $\theta = \pi /4$. The qualitative results for the relative entropy mimic entirely those for the fidelity, and we will thus skip presenting them.

\section{Optimal cheating strategy for Alice}

In this section, we discuss the optimal cheating strategy for Alice, in case she is allowed to perform only single-qubit measurements. In particular, we analyze the effects of noise on the protocol's security in cases when Alice attempts to cheat. Although the requirement of single-qubit measurements might in general pose a significant constraint, for our practical quantum bit commitment scheme it is rather natural. Namely, not only that using today's technology it is not possible to reliably perform large multi-qubit coherent measurements, but in our case Alice would need to have some kind of a stable quantum memory, since Bob sends his qubits sequentially, and at times randomly chosen by him -- precisely the equipment that is not available today and that makes our (practical) commitment scheme possible.

The goal of a cheating Alice is to break one of the two protocol's security requirements: the binding feature. Alice would like to be able to postpone the moment of her commitment, ideally until the opening phase. In order to do that, she has to be able to pass both tests of committing to $0$ and to $1$, and since she is, by the constraints of today's technology, forced to perform her measurements immediately upon receiving the qubits from Bob (during the commitment phase), the only option left is to choose a measurement that would provide her with the best possible inference of qubit states sent by Bob. In other words, the optimal measurement has to secure minimal error when discriminating between the two quantum states. This is a well known problem of ambiguous quantum state discrimination, and the minimal probability of error when discriminating between two general mixed quantum states $\hat{\rho}_0$ and $\hat{\rho}_1$ is given by the famous Helstrom bound \cite{Helstrom}:
\begin{equation}
 P_e(\hat{\rho}_0 , \hat{\rho}_1) = \frac{1}{2} + \frac{1}{2}\mbox{Tr}|p_0\hat{\rho}_0 - p_1 \hat{\rho}_1|,
\end{equation}
where $p_0$ and $p_1 = (1-p_0 )$ are the probabilities of having the state $\hat{\rho}_0$ and $\hat{\rho}_1$, respectively (in our case, $p_0 = p_1 = 1/2$). In the case of pure states and equal a priori probabilities, the Helstrom bound is $P_e = (1 - \sin \theta )/2$ and the optimal observable is given by the orthogonal basis vectors $|\tilde{0}\rangle$ and $|\tilde{1}\rangle$, such that (see for example \cite{State_Estimation}):
\begin{equation}
\begin{aligned}
	|0\rangle & = \cos\alpha |\tilde{0}\rangle + \sin\alpha |\tilde{1}\rangle \\
	|1\rangle & = \cos\beta |\tilde{0}\rangle + \sin\beta |\tilde{1}\rangle\,,
\end{aligned}
\end{equation}
where $\alpha = \pi/4 - \theta/2$ and $\beta = \pi/4 - \theta/2$. In other words, the basis vectors $|\tilde{0}\rangle$ and $|\tilde{1}\rangle$ are in the plane defined by $|0\rangle$ and $|1\rangle$, and share the same bisector with them.

For the value of $\theta = \pi/4$, when the protocol's security is maximal, the optimal observable for a cheating Alice is given by the so-called Breidbart basis \cite{Vlatko-Breidbart}:
\begin{equation}
\begin{aligned}
	|\tilde{0}\rangle & = \cos(\pi/8) |0\rangle - \sin(\pi/8) |1\rangle \\
	|\tilde{1}\rangle & = \sin(\pi/8) |0\rangle + \cos(\pi/8) |1\rangle\,.
\end{aligned}
\end{equation}

\begin{figure}[t!]
	\centering
    \includegraphics[width=0.45\textwidth]{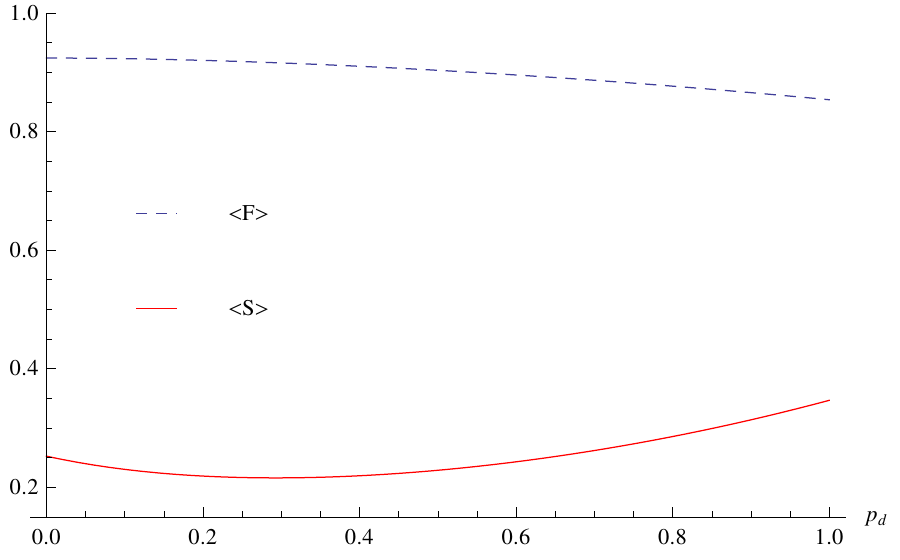}
	\caption{\label{Fid_noiseless_noisycheat_l}(Color online)  Fidelity $\langle F(\mathcal{E} )\rangle$ (dashed, in blue) and relative entropy $\langle S(\mathcal{E} )\rangle$ (full, in red) between a honest strategy without the presence of noise, and the optimal cheating strategy in the presence of white noise, as a function of the noise parameter $p_d$.}
\end{figure}

\begin{figure}[t!]
	\centering
    \includegraphics[width=0.5\textwidth]{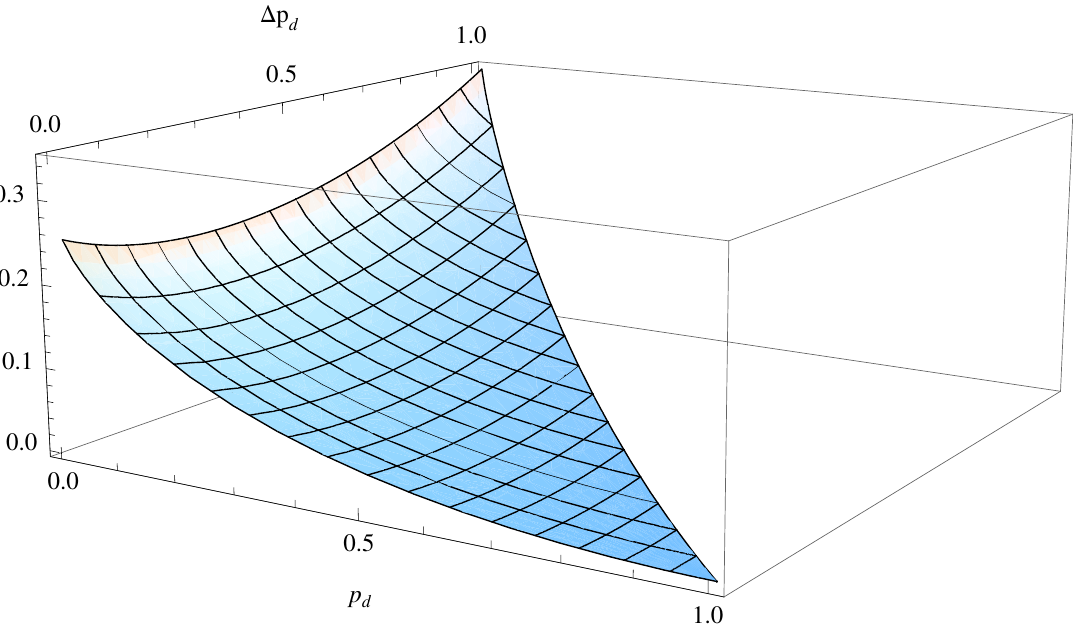}
	\caption{\label{Ent_noiseless_noisycheat_l}(Color online) Introducing a small noise $\Delta p_d$ helps a dishonest party.}
\end{figure}

In order to analyze quantitatively the effects of noise on the above cheating strategy, we compare, in analogy with the previous section, how similar various probability distributions are, using the fidelity and relative entropy as distinguishability measures. In particular, we can consider how different the conditional probabilities obtained by a cheating Alice are from those obtained by the honest one. For simplicity, we start by comparing the results obtained by a cheating party, in the presence of noise, with the results of an honest agent, in the ideal noiseless case \eqref{p_0}, \eqref{p_1}. We will consider the average fidelity $\langle F(\mathcal{E}) \rangle$, given by equations \eqref{fid_noise} and \eqref{fid_noise_4}, where in \eqref{fid_noise_4} instead of $p_0^\mathcal{E}(\ast | \ast)$ and $p_1^\mathcal{E}(\ast | \ast)$, we have the unique cheating probability $p_{ch}^\mathcal{E}(\ast | \ast)$. Analogously, we consider the relative entropy $\langle S(\mathcal{E}) \rangle$. The results for the fidelity and relative entropy are given on Fig.~\ref{Fid_noiseless_noisycheat_l}, respectively (note that in the rest of this section we consider the optimal choice of $\theta = \pi/4$).
We observe the qualitative difference between the behavior of the fidelity and the relative entropy: indeed, one can easily see that the entropy decreases slightly with the introduction of small noises (in fact, the optimal value of added noise is rather significant, being slightly over $0.29$). This behavior can be taken advantage of in the more realistic situation of both parties being subjected to noise. In fact, as Fig.~\ref{Ent_noiseless_noisycheat_l} shows, independently of the channel's noise factor $p_d$, it is always advantageous for a dishonest party to introduce a small extra noise factor $\Delta p_d$, as it decreases the relative entropy between the underlying probability distributions (note that $\Delta p_d \leq 1-p_d$). One can easily prove that the optimal amount of noise $\Delta \tilde{p}_d$ a cheating party should introduce is:
\begin{equation}
\Delta \tilde{p}_d = \bigg(1 - \frac{1}{\sqrt{2}}\bigg)(1 - p_d)\,.
\end{equation}

The above comparative study of the two distinguishability measures shows two rather conflicting results: while according to the fidelity, noise degrades the chances of a cheating party, according to the relative entropy, it is always advantageous to add a little noise in order to increase the chances of a cheating party to go on unnoticed. Unlike the fidelity, the relative entropy gives the probability that a cheating strategy will produce the distribution of measurement results consistent with that produced by an honest party.

It is interesting to analyze the reasons for such behavior of the relative entropy, and why is it not present in the case of the fidelity. For simplicity, we will analyze only the case when an honest strategy is executed in noiseless circumstances, the general case of both honest and cheating agents are subjected to noise is straightforward. First, we note that unlike the standard quantum cryptographic protocols, such as BB84 \cite{bb84} and B92 \cite{b92}, where the only relevant results are those obtained for the cases when the basis of the states sent by Bob and of the observable measured by Alice coincide, here we are interested in the results when the two are not the same. In other words, we are not only interested in how similar the pairs $\left( p_0(\ast | 0) , p_{ch}^\mathcal{E}(\ast | 0) \right)$  and $\left( p_1(\ast | 1) , p_{ch}^\mathcal{E}(\ast | 1) \right)$ are, but also in the distinguishability of the cross terms given by the pairs $\left( p_0(\ast | 1) , p_{ch}^\mathcal{E}(\ast | 1) \right)$ and $\left( p_1(\ast | 0) , p_{ch}^\mathcal{E}(\ast | 1) \right)$ of the probability distributions of the measurement results. And it is precisely the cross terms that make a difference: the honest party probability distributions $p_0(\ast | 1)$ and $p_1(\ast | 0)$ are equal and are actually a uniformly random distribution, while the cheating probability distributions $p_{ch}^\mathcal{E}(\ast | 0)$ and $p_{ch}^\mathcal{E}(\ast | 1)$ are biased, and approach a uniformly random distribution when the noise increases. This means that both the fidelity increase, and the relative entropy decrease, will occur between the pairs of the cross terms for certain range of the noise parameter $p_d$. Nevertheless, these contributions will affect differently the overall average fidelity $\langle F(\mathcal{E}) \rangle$ and the average relative entropy $\langle S(\mathcal{E}) \rangle$: the former will always decrease with $p_d$, while the latter will experience  a decrease for a rather broad range of values of $p_d$. Mathematically, this is just an effect of scaling: fidelity uses a linear scale, and the cross terms are not powerful enough to overcome the behavior of non-cross terms, whereas the relative entropy uses a logarithmic scale, and allows the cross terms to express themselves better.

%\section{ Bob's decision criterion - a more detailed analysis}
\section{ The security of the protocol - a more detailed analysis}
\label{Bob_decision}

In Section \ref{the_protocol} we presented a general description of Bob's decision process: in order to accept Alice's commitment to, say, value $0$, the probability that the statistics $q$, formed by the data communicated by Alice, was obtained by measuring $\hat{C}_0$ must be bigger than a certain threshold value $\alpha$. In addition to that, the probability that the statistics $q$, which passed Bob's test of committing to $0$, was obtained by measuring $\hat{C}_1$ should be smaller than $\beta$ (the protocol is viable). After analyzing the effects of noise and imperfect photon sources and measurement apparatuses, as well as Alice's optimal cheating strategy, we can study in more detail  Bob's criteria for deciding if the results obtained from Alice confirm that she committed to 0, or to 1, or that the results show that she tried to cheat. We can also check the minimum number of photons that need to be measured by Alice for Bob's decisions not to be compromised by Alice's eventual attempt to cheat. 

For simplicity, we will analyze the criterion for deciding if given measurement results confirm that Alice committed to $0$. The total number of measurement outcomes is $n = n(0) + n(1)$, where $n(0)$ is the number of measurements on photons sent in the state $|0\rangle$, and analogously for $n(1)$. Furthermore, $n(0|0)$ is the number of  outcomes $0$, when the state $|0\rangle$ is sent, while $n(1|0)$ is the number of outcomes $1$ (and analogously for $n(\ast |1)$). This way, we have two (conditional) probability distributions:
\begin{equation}
\begin{aligned}
	q(\ast|0) = \{ q(0|0) = \frac{n(0|0)}{n(0)}, q(1|0) = \frac{n(1|0)}{n(0)} \}, \\
    q(\ast|1) = \{ q(0|1) = \frac{n(0|1)}{n(1)}, q(1|1) = \frac{n(1|1)}{n(1)} \}.
\end{aligned}
\end{equation}

\begin{figure*}[t!]
		\centering
	    \includegraphics[width=0.5\textwidth]{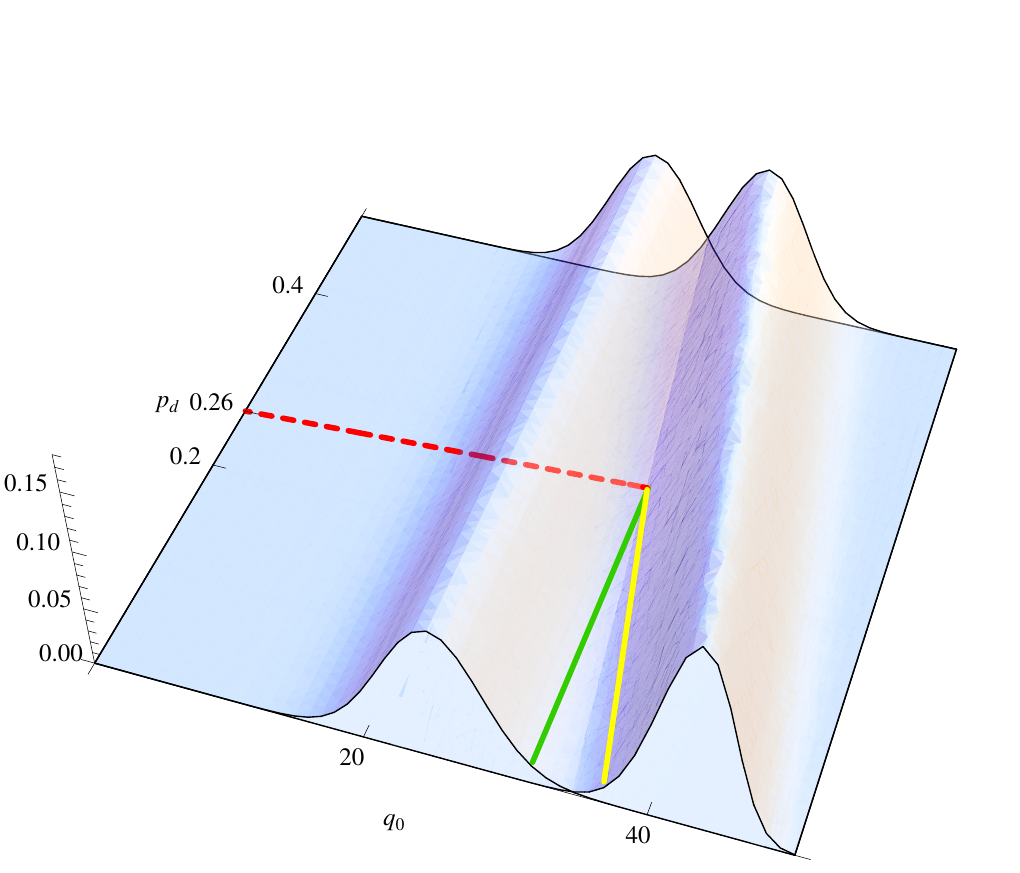}
		\caption{\label{bob_test}(Color online)  $P(q(\ast|1)||p_0(\ast|1))$ -- left bump -- and $P(q(\ast|1)||p_{ch}(\ast|1))$ -- right bump, as functions of a depolarizing coefficient $p_d$ and $q_0= \frac{n(0|1)}{n(0)}$, for $n(0) = 50$. The green (left) full line represents the intersection between the left bump and the $q_0 = \mu_0 + 2\sigma_0$ plane, while the yellow (right) one represents the intersection between the right bump and the $q_0 = \mu_{ch} - 2\sigma_{ch}$ plane, both as functions of $p_d$.}
\end{figure*}

To analyze the protocol's security against Alice's attempt to cheat, we introduce the similar criterion to the one given by (\ref{viable}):
\begin{equation}
	\label{ch_sec}
P(q||p_0) > \alpha \Rightarrow P(q||p_{ch}) < \beta.
\end{equation}
Note that it is more likely to produce statistics that look as if being obtained by committing to $0$ by measuring $\hat{C}_{ch}$ than by measuring $\hat{C}_1$. Therefore, we will only consider the criterion that the statistics $q = \{ 	q(\ast|0), 	q(\ast|1) \}$ were obtained by measuring $\hat{C}_0$, and not by measuring $\hat{C}_{ch}$. Thus, if statistics $q$ pass Bob's test of committing to $0$,
\begin{equation}
	\label{test}
    \begin{aligned}
\!\!	P(q(\ast|0)||p_0(\ast|0)) > \alpha \   \mbox{and} \  	P(q(\ast|1)||p_0(\ast|1)) > \alpha^\prime,
    \end{aligned}
\end{equation}
then for the protocol to be secure (and consequently viable) we require that:
\begin{equation}
	\label{test2}
	\begin{aligned}
\!\!\!\!\!\!\!	P(q(\ast|0)||p_{ch}(\ast|0)) < \beta \ \mbox{and} \  P(q(\ast|1)||p_{ch}(\ast|1)) < \beta^\prime.
    \end{aligned}
\end{equation}
(Note that, in general, the thresholds $\alpha$ and $\alpha^\prime$, as well as $\beta$ and $\beta^\prime$ need not be equal.)

%To estimate (give upper bounds) the above probabilities, we use the following theorem (see {\em Theorem} $4$ from~\cite{vlatko_rel_ent}):
%{\em Theorem:} For any probability distribution $q = \{ k/n, (n-k)/n \}$ ($k\leq n \in \mathbb N$), and any distribution $p$, the probability $P(q||p)$ that $q$ is obtained by the random source $p$ satisfies:
%\begin{equation}
%	P(q||p) \leq e^{-nS(q||p)}.
%\end{equation}

The above probabilities are given by a simple Binomial distribution \footnote{The probability that the distribution $\{p_0 ,1-p_0 \}$ yields statistics $q$ (that is, the probability that $\{p_0 ,1-p_0 \}$ yields $n_0$ 0s and $(n-n_0)$ 1s) is given by $p_0^{n_0} (1-p_0)^{(n-n_0)}$ times the number of possible sequences of $n_0$ 0s and $(n-n_0)$ 1s, i.e. $\left( \begin{array}{c} n \\ n_0 \end{array} \right)$.} $\mathcal{B}(n_0;n,p_0)$ (for simplicity, here we define $n = n(0)$, $n_0 = n(0|1)$ and $p_0 = p_0(0|1)$):
\begin{equation}
	\label{binomial}
	P(q(\ast|1)||p_0(\ast|1)) = \left( \begin{array}{c} n \\ n_0 \end{array} \right) p_0^{n_0} \left( 1 - p_0 \right)^{(n- n_0)},
\end{equation}
and analogously for other cases. For all practical purposes the above Binomial distribution will behave as a Normal distribution $\mathcal{N}(x;\mu_0,\sigma_0)$, with its mean $\mu_0$ and standard deviation $\sigma_0$. Thus, we can set parameter $\alpha$ to be $\alpha_{2\sigma} = \Phi (\mu_0 +2\sigma_0;\mu_0,\sigma_0)$, where $\Phi (x;\mu_0,\sigma_0)$ is the cumulative distribution function of the normal distribution $\mathcal{N}(x;\mu_0,\sigma_0)$. For such $\alpha$ the statistics obtained by measuring $\hat{C}_0$ will pass the test of committing to $0$ in about $97.7\%$ of the cases. Analogously, the distribution defining the second probability in (\ref{test2}) is given by $\mu_{ch}$ and $\sigma_{ch}$, and we can define  $\beta$ to be $\beta_{2\sigma} = \Phi (\mu_{ch} - 2\sigma_{ch}; \mu_{ch},\sigma_{ch})$, for which the statistics obtained by measuring $\hat{C}_{ch}$ will lead to a successful cheat in only $2.3\%$ of the cases.

It turns out that the second conditions in both (\ref{test}) and (\ref{test2}) are, albeit qualitatively the same, quantitatively stronger than the first pair of conditions. In Fig. \ref{bob_test} we plot $P(q(\ast|1)||p_0(\ast|1))$ -- left bump -- and $P(q(\ast|1)||p_{ch}(\ast|1))$ -- right bump, as functions of a depolarizing coefficient $p_d$ and $q_0= \frac{n(0|1)}{n(0)}$, for $n(0) = 50$ (i.e. the total number of measurements is about $n = 100$). The second condition in (\ref{test2}) now translates to
\begin{equation}
	\label{cond_alpha_beta}
\mu_0 +2\sigma_0 \leq \mu_{ch} -2\sigma_{ch}.
\end{equation}
The latter is satisfied as long as the depolarizing coefficient $p_d$ is smaller than about $26\%$, which is represented by the intersection of the two full lines in Fig. \ref{bob_test}.
%The plot reveals that a statistics $q(*|0)$ that passes condition \ref{test} can only be obtained, up to negligible probability, by measuring $\hat{C}_0$. It is impossible for Alice to obtain a statistics $q(*|0)$ by measuring $\hat{C}_{ch}$ that passes the same test since the two bump functions barely intersect. 
%For this same reason, the upper bounds given in the previous \emph{Theorem} can be used in place of the actual probabilities $P(q(\ast|0)||p_0(\ast|0))$ and $P(q(\ast|0)||p_{ch}(\ast|0))$ in condition (\ref{test}).

\begin{table}
\begin{tabular}{ | x{2.5cm} | x{2.5cm} | c | }
  \hline                       
  $\alpha$ & $\beta$ & $p_d^\ast$ \\
\hline \hline
  $\alpha_{2\sigma}  =97.7\% $ & $\beta_{2\sigma} =2.3\%$ & \mbox{ } 0.26 \mbox{ } \\
\hline
  $ \alpha_{3\sigma} = 99.86\% $ & $\beta_{\sigma} =15.8\% $ & 0.23 \\
\hline
  $ \alpha_{3\sigma} = 99.86\% $ & $\beta_{2\sigma} = 2.3\%$ & 0.09 \\
\hline
  $ \alpha_{2\sigma} = 97.7\% $ & $\beta_{\sigma} =15.8\% $ & 0.42 \\
  \hline  
\end{tabular}
\caption{Maximum value of $p_d$ as a function of the security parameters $\alpha$ and $\beta$.}
	\label{table}
\end{table}

We see that already for $n = 100$ we obtain statistics such that the probability to cheat becomes negligible for realistic amounts of noise. Note that this number is smaller than the number of photons needed to be measured in standard quantum key distributions, where a number of results are used not for establishing a secret key, but ``wasted'' for checking if the communication between Alice and Bob was eavesdropped.

The conditions determining the probabilities $\alpha$ and $\beta$ in (\ref{test2}), given by (\ref{cond_alpha_beta}), can be either strengthened or loosened according to one's needs. One can, for instance, increase the viability of the protocol by setting $\alpha$ to be $\alpha_{3\sigma} = \Phi (\mu_0 + 3\sigma_0; \mu_0,\sigma_0)$, in which case an honest party would successfully pass Bob's test in $99.86\%$ of the cases. On the other hand, one could decrease the security parameter $\beta$ to $\beta_{\sigma} = \Phi (\mu_{ch} - \sigma_{ch}; \mu_{ch},\sigma_{ch})$, which would still allow Bob to spot cheating in about $84.2\%$ of the cases. In Table \ref{table}, we present various possibilities for $p_d^\ast$, the maximum value of $p_d$ for which conditions (\ref{test}) and (\ref{test2}) are satisfied, as functions of the security parameters $\alpha$ and $\beta$ (i.e., the conditions analogous to (\ref{cond_alpha_beta})).

%We haven't yet fully described a way for Bob to decide whether or not to accept a certain commitment by Alice, based on a given set of statistics $q$. We will do so here, and simultaneously show that the number of qubits required for a decent statistical sample is low enough as to guarantee the practicality of the protocol. First, recall the following results: suppose we're given a sequence $b_1b_2 \dots b_n$ of bits, and consider the inherited probability distribution $Q$ given by
%\[
%\{ \frac{\text{number of 0's in the sequence}}{n} ,\frac{\text{number of 1's in the sequence}}{n}\} .
%\]
%Let $B$ be a binomial distribution arising from a sequence of $n$ Bernoulli trials. Then the probability $P$ of obtaining a sequence of bits whose inherited probability distribution is $Q$ satisfies
%\[
%P\leq e^{-n S(Q||B)} .
%\]
%In other words, the probability of drawing a sample according to $B$, but whose inherited probability distribution is $Q$ decreases exponentially with $n$ (unless the parameter of the Bernoulli trial $Q$ matches that of the Bernoulli trial underlying distribution $B$).

%With this in mind, it is now quite simple to state how likely is Bob to convince himself that Alice committed to a certain bit value $c$ when she actually didn't.

\section{(Im)perfect non-demolition measurements and noisy/bounded memory}

\begin{figure*}[t!]
		\centering
	    \includegraphics[width=0.5\textwidth]{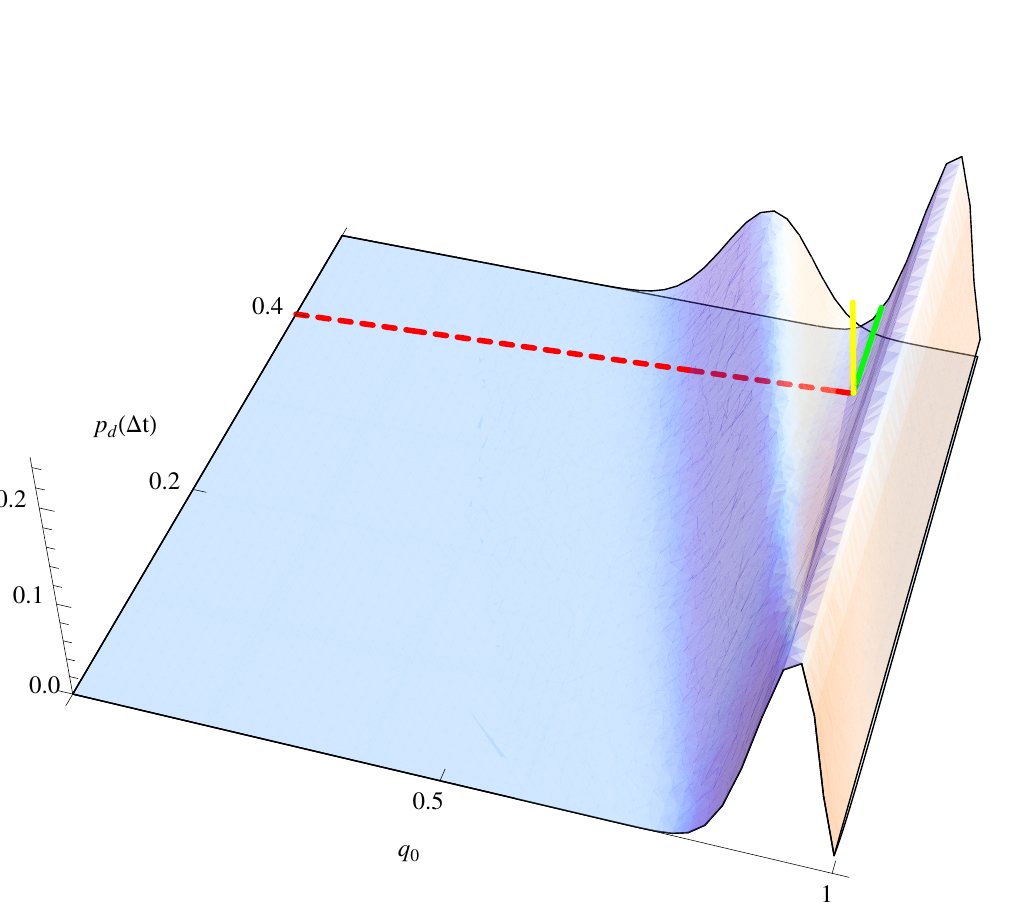}
		\caption{\label{bob_with_memory}(Color online)  $P(q(\ast|0)||p_0(\ast|0))$ -- right bump -- and $P(q(\ast|0)||p_0^{\Delta t}(\ast|0))$ -- left bump, as a function of $p_d (\Delta t)$ and $q_0= \frac{n(0|1)}{n(0)}$, for a fixed value of $p_d =0.15$ and $n(0) = 50$. The green (right) full line represents the intersection between the right bump and the $q_0 = \mu_0 - 2\sigma_0$ plane, while the yellow (left) one represents the intersection between the left bump and the $q_0 = \mu_0^{\Delta t} + 2\sigma_0^{\Delta t}$ plane, both as functions of $p_d (\Delta t)$.}
\end{figure*}

In this section we analyze the protocol's security under the more realistic assumptions of using finite efficiency non-demolition measurements and a noisy/bounded memory.

First, we discuss the use of non-demolition measurements. The ideal non-demolition photon measurement would allow Alice to obtain the photon arrival times without actually destroying them, thus permitting her to keep the qubits in a (noisy) memory and postpone her commitment. Typically, such non-demolition measurements would alter the photon's state of polarization, a contribution which is equivalent to one coming from an imperfect memory and with which we'll deal later on. In case of a finite efficiency, say $p_{nd} < 1$, a certain fraction of arrived photons will not be possible to store. If Alice's equipment simply absorbs such photons, and does not allow measuring the photon polarizations, Bob can detect such cheating attempt by comparing the expected and presented number of results (Bob  knows the specifications of the set-up, in particular the rate of photon emission, the absorption coefficient of the environment, and the detectors' efficiencies, and can therefore estimate the expected number $n$ of results provided by Alice). In case she has an apparatus that can measure the polarization of the absorbed photons, Alice's best strategy is to perform the polarization measurement of the cheating observable $\hat{C}_{ch}$ on the said fraction of $(1-p_{nd})n$ photons. But this is equivalent to the case of performing an ideal non-demolition measurement, and having a bounded noisy memory, such that only the fraction of $\nu = p_{nd}$ of results are obtained by the stored photons, while the rest are obtained by measuring $\hat{C}_{ch}$. Therefore, in the rest of the section we assume the ideal case of $p_{nd} = 1$.

%If $p_{nd}$ is too low, then Bob can detect such cheating by comparing the expected and presented number of results (he knows the specifications of the set-up, in particular the rate of photon emission, the absorption coefficient of the environment, and the detectors' efficiencies). If, on the other hand, $p_{nd}$ is sufficiently high, then a cheating Alice's best bet is to perform the polarization measurement of the cheating observable $\hat{C}_{ch}$ on the said fraction of photons. This case is then equivalent to the case of ideal non-demolition measurements in the presence of bounded quantum memory which will be analyzed below. In other words, the non-demolition measurements should be of high efficiency, and the analysis becomes the same as for the ideal case of $p_{nd} = 1$.

Regarding quantum memories, we will first again discuss the best case scenario for a cheating Alice: unrestricted amount of noisy quantum storage. Storing qubits in a memory for a certain ``delay time'' $\Delta t$ allows Alice to postpone her commitment and thus break the binding security condition. Since the memory is not ideal, during that time the photon's polarization states will further decohere with the environment, hence decreasing Alice's probability to pass Bob's test (\ref{test}). We model the noise by a depolarizing quantum channel given by $p_d(\Delta t)$. This way, when measuring, say $\hat{C}_0$, the cheating a priori conditional probability distributions $p_0^{\Delta t}(\ast|\ast)$, given by $p_d+p_d(\Delta t)$, will differ from those expected by Bob, given by only $p_d$. As in the previous section, the threshold value for $p_d(\Delta t)$, for which cheating is not possible, is given by (Bob's decision criterion)
\begin{equation}
	\label{test0}
    \begin{aligned}
\!\!	P(q(\ast|0)||p_0(\ast|0)) > \alpha \   \mbox{and} \  	P(q(\ast|1)||p_0(\ast|1)) > \alpha ,
    \end{aligned}
\end{equation}
and (security criterion)
\begin{equation}
	\label{test02}
	\begin{aligned}
\!\!\!\!\!\!\!	P(q(\ast|0)||p_0^{\Delta t}(\ast|0)) < \beta^{\Delta t} \ \mbox{and} \  P(q(\ast|1)||p_0^{\Delta t}(\ast|1)) < \beta^{\Delta t}.
    \end{aligned}
\end{equation}
Note that $\alpha$ is given in terms of $\mu$ and $\sigma$ of the ``honest'' conditional probability distributions $p_0(\ast|\ast)$, while $\beta^{\Delta t}$ is obtained for $\mu^{\Delta t}$ and $\sigma^{\Delta t}$ of the ``cheating'' conditional probability distributions $p_0^{\Delta t}(\ast|\ast)$.

Again, we analyze the case of committing to $0$, and for simplicity we choose only the $q(\ast|0)$ results. Further, in analogy with the previous section, we (re)define $n = n(0)$, $n_0 = n(0|0)$, $p_0 = p_0(0|0)$ and $q(\ast|0) = \{ q = n_0 /n, 1-q = (1-n_0)/n \}$. This change of notation is due to the fact that $p_0^{\Delta t}(\ast|1) =p_0 (\ast|1)$ is uniformly random, and independent of $\Delta t$. In Fig. \ref{bob_with_memory}, we plot $P(q(\ast|0)||p_0(\ast|0))$ -- right bump -- and $P(q(\ast|0)||p_0^{\Delta t}(\ast|0))$ -- left bump, as functions of $p_d(\Delta t)$, for a fixed value of $p_d =0.15$. For higher (lower) values of $p_d$, the corresponding plot is a simple translation to the left (right), along with a rescaling.

Finding the threshold value $p_d^\ast (\Delta t)$ for $p_d(\Delta t)$ (i.e. for how long can Alice postpone her commitment), as a function of $p_d$, is now equivalent to solving
\begin{equation}
\mu_0^{\Delta t}+2\sigma_0^{\Delta t} \leq \mu_0 -2\sigma_0,
\end{equation}
just as in the previous section (see Eq. (\ref{cond_alpha_beta})). Again, we can vary the security parameters $\alpha^{\Delta t}$ and $\beta^{\Delta t}$. Various possibilities are presented in Table \ref{table2}.

\begin{table}
\begin{tabular}{ | x{2.5cm} | x{2.5cm} | c | }
  \hline                       
  $\alpha^{\Delta t}$ & $\beta^{\Delta t}$ & $p_d^\ast (\Delta t)$ \\
\hline \hline
  $\alpha_{2\sigma}  =97.7\% $ & $\beta_{2\sigma} =2.3\%$ & \mbox{ } \mbox{ } 0.4 \mbox{ } \mbox{ } \\
\hline
  $ \alpha_{3\sigma} = 99.86\% $ & $\beta_{\sigma} =15.8\% $ & 0.35 \\
\hline
  $ \alpha_{3\sigma} = 99.86\% $ & $\beta_{2\sigma} = 2.3\%$ & 0.49 \\
\hline
  $ \alpha_{2\sigma} = 97.7\% $ & $\beta_{\sigma} =15.8\% $ & 0.26 \\
  \hline  
\end{tabular}
\caption{Maximum value of $p_d (\Delta t)$ as a function of the security parameters $\alpha$ and $\beta$.}
	\label{table2}
\end{table}

Finally, we briefly discuss the general case of bounded and noisy quantum memories. Let $\nu$ be the fraction of the total results $n$ which were obtained by measuring the photons stored in a noisy quantum memory. The optimal strategy for a cheating Alice would then be to measure $\hat{C}_0$ on $\nu n$ photons, and the cheating observable $\hat{C}_{ch}$ on the rest of $(1-\nu)n$ photons that she cannot store. Due to the law of large numbers, the a priori probability of such cheating strategy would be
\begin{equation}
\bar{p}_0(\ast|0) = \nu p_0(\ast|0) + (1-\nu)p_{ch}(\ast|0),
\end{equation}
and analogously for $\bar{p}_0(\ast|0)$. The security criterion would then be:
\begin{equation}
	\label{test02}
	\begin{aligned}
\!\!\!\!\!\!\!	P(q(\ast|0)||\bar{p}_0(\ast|0)) < \bar{\beta} \ \mbox{and} \  P(q(\ast|1)||\bar{p}_0(\ast|1)) < \bar{\beta}^\prime ,
    \end{aligned}
\end{equation}
which can be easily analyzed using the previous results regarding the cheating observable and the noisy memory.

\section{Non-optical detector errors}

In this section, we briefly discuss the effects of non-optical errors, caused by the imperfect single photon sources, transmission losses and imperfect detectors (finite efficiency and dark counts). As the causes of these errors are basis-independent, they will manifest equally as in the case of standard quantum cryptography, when the state $|b\rangle$ sent by Bob and the observable $\hat{C}_c$ measured by Alice coincide, $c=b$. In Appendix \ref{qber}, following \cite{Gisin-review}, page $166$, we present the explicit expressions for non-optical quantum bit corrections $\delta p_c(\ast | b)$, for $c=b$, to the total probabilities $\tilde{p}_c(\ast | b) = p_c(\ast | b) + \delta p_c(\ast | b)$, where $p_c(\ast | b)$ represent optical contribution discussed in Section \ref{Optical_noise}.

As in the case of calculating the non-optical part of QBER, so in cases of measuring a ``wrong'' observable, when $c \neq b$, we neglect all less probable cases and calculate the error due to dark counts only when no photons arrived at the measuring apparatus (that consists, among other things, of two detectors $D_0$ and $D_1$, corresponding to two possible outcomes $0$ and $1$, respectively). Therefore, we can safely estimate that the number of dark counts in both detectors is equal. In the case of $|\langle 0|1\rangle| = \cos \pi/4 = 1/\sqrt{2}$, when the protocol's security is optimal, the error due to dark counts is thus zero.

\section{Conclusions}

We presented a two-state practical quantum bit commitment protocol. We discussed the effects of both optical and non-optical noise. In the latter case, we showed that finite detector efficiency, dark counts, imperfect single-photon sources and transmission losses have essentially the same effects as in the case of standard quantum cryptography. To quantitatively analyze the effects of the optical part of the noise, we used the fidelity and the relative entropy, two information-theoretic measures of probability distribution distinguishability. As a corollary to our study, using the two distinguishability measures only, we obtained the well-known result that the optimal value of the angle $\theta$ between the two quantum states used in the protocol is $\theta = \pi/4$. We also showed a somewhat counter-intuitive result that adding a certain amount of white noise can always help a cheating Alice to postpone her commitment until the opening phase. This effect is a result of the comparison of the results of measurements in cases when the measurement basis do not coincide with the basis from which the state is sent. Although it can be seen by looking at the behavior of both the fidelity and the relative entropy, when averaging over all the possible cases, only the expected relative entropy retains the signature of this effect. Finally, we analyzed the protocol's security when Alice has access to (im)perfect non-demolition measurements and noisy/bounded memories, and showed that the protocol is robust against such possible attacks.

%%%%%%%%%%%%%%%%%%%% Appendix %%%%%%%%%%%%%%%%%%%%%%%%%%%%%%%%%%%%%%%%%%%%%%%%%%%%%%%%%%%%%%%%%%%%%%%%%%%%%%%%%%%%%%%%%%%%%%%%%%%%%%%%%%%%%%%%%%%%%%%%%%%%%%%%%%%%%%%%%

\appendix\section{}
\label{bit_commitment_applications}

In this Appendix, we present a simple example of the application of bit commitment to authentication protocols based on zero-knowledge proof systems. Suppose Alice wants to authenticate herself to Bob, by proving to Bob that she knows a solution to a difficult mathematical problem, without actually revealing the solution (thus the name zero-knowledge proof). This requirement is crucial: the knowledge Alice has is unique to her (the problem is difficult, so others cannot solve it {\em in real time}), and is used as a mean of identification. If she discloses it to Bob, he can in the future falsely present himself as Alice, which she would like to prevent.

Consider the following mathematical problem (the so-called {\em coloring problem}): given a graph $G = (V,E)$, where $V$ is the set of vertices and $E \subseteq V \times V$ the set of edges, and three colors $\{ R,Y,B \}$ -- red, yellow and blue -- find a coloring $C: V \rightarrow \{ R,Y,B \}$, such that no two adjacent vertices have the same color, $(u,v) \in E \Rightarrow C(u) \neq C(v)$. This is known to be a hard problem, in fact it is an NP-complete problem (see for example \cite{colouring}, page $1019$): if only a graph is given, finding a proper coloring using today's best algorithms requires exponential time, with respect to the graphs' complexity. Therefore, it is for all practical purposes safe to assume that Alice is the only person who knows a coloring, and therefore she can use it as her personal identifier.

The way to prove to Bob that she indeed knows the coloring $C$, without actually revealing this information, is the following. The protocol is probabilistic, consisting of $n$ steps, such that the probability that Alice cheats (convinces Bob she knows the coloring, without actually knowing $C$) approaches to zero exponentially fast, with respect to the number of steps $n$. Each step consists of three consecutive parts:
\begin{enumerate}
\item Alice randomly chooses a permutation $\pi : \{ R,Y,B \} \rightarrow \{ R,Y,B \}$, sets a new coloring $C^\prime = \pi \circ C$, and {\bf commits} to it: writes down on a piece of paper the colors, according to new coloring $C^\prime$, of all the vertices of a (publicly known) graph $G$, {\em locks} it in a secure ``safe'', keeps the key with her, and gives the ``safe'' to Bob. She can do so by committing to a string of $2N$ bits, where $N$ is the number of vertices of $G$: assuming Alice and Bob agreed prior to the protocol on a particular enumeration of the graph's vertices, each $i$-th pair of bits, with $i = 1, \ldots N$, defines the color of the $i$-th vertex  (obviously, this is not an optimal encryption); this way, each bit is locked in a different ``safe'', for which a different key is produced.
\item Bob chooses an edge $(u.v)$ and challenges Alice to show him their respective colors.
\item Alice {\bf opens} the values of the bit pairs corresponding to the vertex $u$ and the vertex $v$ (gives the keys for the corresponding bits), thus disclosing to Bob the colors $C^\prime(u)$ and $C^\prime(v)$. If they are the same, Alice failed to pass the test and Bob terminates the procedure. Otherwise, they repeat the procedure until Bob is satisfied.
\end{enumerate}
Obviously, Alice can pass the above test only if she indeed knows the coloring $C$ of the graph $G$. Otherwise, she can only try to partially color the graph (properly, so that the adjacent vertices have different colors), hoping that Bob will not choose vertices that she colored with the same color. If the probability to pass the test in a single step of the protocol is $p < 1$, then the probability to pass the test goes to zero exponentially fast with the number of steps $n$ of the protocol, as $1-p^n$. Note that although in each step of the protocol Bob learns a coloring of one pair of vertices, after $n$ steps he still did not learn the coloring of $n$ vertices, as in each step Alice chooses different coloring $C^\prime = \pi \circ C$, given by a permutation $\pi$, unknown to Bob.

Note the essential importance that Alice's commitment is binding -- otherwise, she could, upon learning Bob's choice of vertices $u$ and $v$, change her commitment and choose the two colors to be different. Also, it is important that the protocol is concealing -- otherwise, Bob would be able to learn a coloring of graph $G$, and thus, in the future, impersonate Alice.

This concept of a cryptographic commitment can be traced back to the early 80's, with the works of Shamir, Rivest and Adleman \cite{RSA81}, along with those of Blum \cite{Blum82} and finally of Even \cite{Even82} where the concept was first named. Nowadays, it has found its way into several protocols of many diverse natures, such as e-voting protocols \cite{CarlosRibeiro11}, the TESLA authentication protocol \cite{TESLA05}, and the Schnorr protocol \cite{Schnorr90} on which part of Microsoft's U-prove system is based \cite{UPROVE10}.

\section{}
\label{qber}  %multi_photon

In this Appendix, we evaluate the non-optical contributions $\delta p_c(\ast|b)$, for $c=b$. As noted when discussing the depolarizing channel, the two out of four probabilities are nothing but the quantum bit error rate, $\mbox{QBER} = \delta p_0(1|0) = \delta p_1(0|1)$, while the other two are then straightforward to obtain, $\delta p_0(0|0) = - \delta p_0(1|0)$ and $\delta p_1(1|1) = - \delta p_1(0|1)$ (note that, by definition, $\tilde{p}_0(0|0) + \tilde{p}_0(1|0) = 1$ and $p_0(0|0) + p_0(1|0) = 1$). For example, let Alice measure $\hat{C}_0$. Then, we are interested in cases when result $1$ is obtained for qubits in state $|0\rangle$. Let $N_{\text{tot}}$ be the total number of qubits received in the state $|0\rangle$, and let $N_{\text{\text{wrong}}}$ be the number of qubits received in the state $|0\rangle$ for which the wrong result $1$ is obtained, during the time interval $T$. Then, the QBER is given by:
\begin{equation}
	\label{QBER}
\mbox{QBER} = \frac{N_{\text{wrong}}}{N_{\text{tot}}} = \frac{\frac{N_{\text{wrong}}}{T}}{\frac{N_{\text{tot}}}{T}} = \frac{R_{\text{error}}}{R_{\text{tot}}}\,.
\end{equation}
Here $R_{\text{tot}} = N_{\text{tot}}/T$ and $R_{\text{error}} = N_{\text{wrong}}/T$ are the total and the error rates, respectively, for the qubits received in the state $|0\rangle$. If $R_{\text{raw}}$ is the overall source rate, including both $|0\rangle$ and $|1\rangle$ states, then
\begin{equation}
R_{\text{tot}} = \frac{1}{2} R_{\text{raw}},
\end{equation}
since Bob sends on average equal number of $|0\rangle$ and $|1\rangle$ states. The total rate (number/time) of qubits sent in either $|0\rangle$ or $|1\rangle$ state (given by $f_{\text{rep}}\mu$), that were not absorbed and that managed to arrive to detectors (given by $t_{\text{link}}$) and were detected (given by $\eta$):
\begin{equation}
R_{\text{raw}} = f_{\text{rep}}\mu t_{\text{link}}\eta\,.
\end{equation}
Here, $f_{\text{rep}}$ is the pulse rate (the number of ``attempts'' to send a photon, per time), and $\mu$ is the mean number of photons per pulse. Thus, $f_{\text{rep}}\mu$ is the number of photons sent, in the unit of time. The probability that a sent photon arrives to a detector is $t_{\text{link}} \sim 10^{-\alpha L}$ ($\alpha$ is the absorption coefficient, and $L$ is the transmission distance, i.e. length of an optical cable). Finally, the detector efficiency $\eta$ is the probability that a photon that arrived to a detector is actually detected. Note that $\mu \sim  0.1 << 1$: we ignore the low probable cases of sending two or more photons per pulse.

In general, $R_{\text{error}} = R_{\text{opt}} + R_{\text{det}}$, but here we are only interested in the error arising due to dark counts. We have
\begin{equation}
R_{\text{det}} = \frac{1}{2} \bigg(\frac{1}{2}f_{\text{rep}}\bigg) (1 - \mu t_{\text{link}}\eta) p_{\text{dark}} \approx \frac{1}{4} f_{\text{rep}} p_{\text{dark}}\,.
\end{equation}
Here, $\frac{1}{2}$ is the probability that a wrong detector will click; $(\frac{1}{2}f_{\text{rep}})$ is the rate of photons sent in the ``right'' state (in our case $|0\rangle$); $(1 - \mu t_{\text{link}}\eta)$ is the probability that a photon is not detected (when dark counts are relevant!). Note that the probability that a photon arrives to detectors and is efficiently detected is small, $\mu t_{\text{link}}\eta << 1$. Also, we neglect the less probable cases of ``right'' dark counts, when a photon does not arrive to detectors.

Thus, we get:
\begin{equation}
\begin{aligned}
	\delta p_0(1|0) & = \delta p_1(0|1) = \frac{p_{\text{dark}}}{2 \mu t_{\text{link}} \eta} \\
	\delta p_0(0|0) & = \delta p_1(1|1) = - \frac{p_{\text{dark}}}{2 \mu t_{\text{link}} \eta}\,.
\end{aligned}
\end{equation}

%%%%%%%%%%%%%%%%%%%%%%%%%%%%%%%%%%%
\begin{acknowledgments}
RL thanks Funda\c{c}\~{a}o para a Ci\^{e}ncia e a Tecnologia - FCT the support through the PhD Grant SFRH/BD/79571/2011.

AA thanks the FCT the support through the PhD Grant SFRH/BD/79482/2011.

AA, PA and AP thank the project PTDC/EEA-TEL/103402/2008 (QuantPrivTel); the FCT and the Instituto de Telecomunica\c{c}\~{o}es - IT under the PEst-OE/EEI/LA0008/2013 program, project `P-Quantum', and the Conselho de Reitores das Universidades Portuguesas (CRUP) project `A\c{c}\~{a}o Integrada E 91/12'.

PM and NP thank the project of SQIG at IT, funded by FCT and EU FEDER projects PTDC/EIA/67661/2006 (QSec), PTDC/EEATEL/103402/2008 (QuantPrivTel), FCT PEst-OE/EEI/LA0008/2013 and IT Projects QuantTel and `P-Quantum', as well as Network of Excellence, Euro-NF.

PM also thanks ComFormCrypt PTDC/EIACCO/113033/2009.
\end{acknowledgments}

%---------------------------------------------------------

\end{document}